\documentclass[twocolumn]{aastex7}
\usepackage{amsmath}
\usepackage{empheq}
\usepackage[bb=dsserif]{mathalpha}
\usepackage{bm}

\begin{document}

\title{Effects of Super-rotating Jets on Phase-Resolved Transmission Spectra at High Spectral Resolution}

\author[orcid=0000-0002-6980-052X]{Hayley Beltz}
\affiliation{Department of Physics and Astronomy, University of Kansas}
\email[show]{hbeltz@ku.edu}  

\author[orcid=0000-0002-2454-768X]{Arjun B. Savel} 
\affiliation{Department of Astronomy, University of Maryland, College Park}
\email{asavel@umd.edu}



\begin{abstract}

Hot Jupiter atmospheric circulation can be shaped by a range of processes and planetary parameters, including rotation rate, irradiation gradients from tidal synchronization, magnetic effects, gravity, and clouds. Observations of exoplanetary spectral Doppler shifts as a function of phase, made possible with high-resolution spectrographs, can now probe the effects of exoplanetary winds at several points throughout transit. However, these measurements are difficult to interpret, given the variety of relevant atmospheric mechanisms. In this work, we generate high-resolution transmission spectra from two sets of circulation models: self-consistent dynamical models and idealized models that isolate circulation components. We identify how the strength of the jet --- or the absence of one --- alters the resultant net Doppler shifts. We find that the jet strength most prominently affects Doppler shifts during ingress and egress. During mid-transit, the strength of the jet linearly influences the Doppler shifts’ slope; however, this effect is generally secondary to the overall blueshifting from planetary rotation. The slope induced by equatorial jets contrasts with the effect of pure day--night flows, which tend to add constant offsets to Doppler shifts during transit. Our modeling shows that jets also impact the cross-correlation function width; faster jets increase velocity dispersion across the limbs. We complement our simulations with semi-analytical arguments indicating that first-order changes of the cross-correlation function centroid and width (but not amplitude) probe thermal and velocity asymmetries. This work provides recommendations for interpreting net Doppler shifts in transmission and connecting these shifts with exoplanets’ atmospheric circulation.
\end{abstract}



\section{Introduction}\label{sec: intro} 
The advent of high-resolution cross-correlation spectroscopy (HRCCS), with its unprecedented resolving power, has dramatically changed the landscape of exoplanet atmospheres studies. This increase in resolution allows for the measurement of atmospheric winds, making it a unique tool for probing circulation dynamics of exoplanets \citep{Snellen2010}.\footnote{ For recent reviews on HRCCS use in exoplanet atmospheres, see \cite{Birbky2018, Snellen2025}. } Planetary atmospheres are multidimensional objects, and this multidimensionality persists in HRCSS.  Previous work \citep{Flowers2019} found that for transmission  spectroscopy,  the inherent broadening from the 3D models provided an excellent match to the line-of-sight velocity profiles without the need to invoke additional broadening or scaling, as required by 1D models, indicating that 3D effects are present in the data. High resolution spectrographs such as ESPRESSO \citep{Espressopepe2014}, CRIRES$^+$ \citep{CRIRES+2014}, PEPSI \citep{Pepsi2025}, MAROON-X \citep{Maroonx2018}, NIRPS \citep{NIRPS2025}, and IGRINS \citep{IGRINS2014} can make high-precision measurements of the atmosphere at multiple points throughout orbit that can then be linked to  exoplanetary wind speed. This technique of phase-resolved Doppler shifts paints a complex picture of a planet's atmospheric dynamics \citep{Komacek2020, Borsa2021,Pino2022,Prinoth2023,Wardenier2024,simonnin2025time}. 

However, interpreting these net Doppler shifts has proven to be difficult, due to the multidimensional nature of the atmospheres as well as the multiple concurrent physical processes at work. For example, \cite{Ehrenreich2020} measured net Doppler shifts for Fe I during transit for the ultrahot Jupiter (UHJ)  WASP-76b. The authors suggest that these data were representative of Fe I condensation on the nightside of the planet. However, sophisticated 3D models had difficulty matching the observed trend and magnitude, with proposed explanations spanning rainout, temperature differences \citep{Wardenier2021}, and  optically thick clouds \citep{2022Savel}. 
\cite{Beltz2023}  additionally modeled the planet with 3D models under various magnetic assumptions and found that a simplified magnetohydrodynamical (MHD) prescription approach was able to reproduce the trends seen in \cite{Ehrenreich2020}, but not the magnitude of the Doppler shifts.

One can also consider probing the atmosphere vertically with this technique. Theoretical expectations from General Circulation Models (GCMs) suggest that the presence of magnetic drag will alter the net Doppler shifts as a function of pressure  \citep{Kempton2012}. Critically, they found that the expected net Doppler shifts of their drag-free model decreased with increasing pressure and the reverse for their active drag model. The ``active drag'' approach, sometimes referred to as kinematic MHD, is a numerical approximation of magnetic effects that uses a locally calculated drag timescale to simulate the effect of the Lorentz force felt by charged particles as they are blown across magnetic field lines \citep{Perna2010magdrag}. This approach is used in this work; see Section \ref{sec: GCMmethods} for more detail.  \cite{KesseliBeltz2024} applied this concept to the same ESPRESSO data as \cite{Ehrenreich2020} and broke up iron lines into three different bands, each corresponding to different strengths of lines. These different line strengths probed different pressure regions; the strongest iron lines probed the highest altitude while the weak lines probed deeper into the atmosphere. They then compared the corresponding Doppler shifts to multiple GCMs, coming to the conclusion that the model with the active drag magnetic prescription  was preferred. Recently, \cite{Seidel2025} also explored the vertical atmospheric structure of the UHJ WASP-121b by utilizing three different opacity sources that probe different pressure regions. They interpret their net Doppler shifts to be consistent with an atmosphere where a jet lies above a region of day-to-night flow and below a region of outward flow escaping the atmosphere. This is in tension with 3D GCMs \citep{PerezBecker2013,Showman:2020rev}, which predict jets to extend deep into the atmosphere with day-to-night flow and escaping outflow higher in the atmosphere. To our knowledge, no existing 3D atmospheric model can reproduce this structure. 

Many of the published theoretical predictions of phase-resolved Doppler shifts are narrow in scope, focusing on a specific planet or a specific absorber \citep{Wardenier2021,2022Savel,Beltz2023, Beltz2024, Wardenier2024}. This presents an opportunity to study the broad range of atmospheric dynamics at work in HJs/UHJs by leveraging phase-resolved Doppler shift observations from these state-of-the-art ground-based facilities. With this in mind, we set out to explore how a jet may influence net Doppler shifts in transmission spectra.

\subsection{Equatorial Jets in Hot Gas Giant Atmospheres}

The eastward superrotating equatorial jet with speeds on order of a few km/s is robustly predicted across multiple GCMs of hot and ultrahot Jupiters \citep{Showman2002, Heng2011, 2012RauscherGCM, Mayne2014, 2013DobbsDixon, parmentier2018, Deitrick2022THOR+HELIOS} and displays very little hysteresis/dependency on initial conditions \citep{Komacek2025hysteresis}.\footnote{This assumes that the model does not have excessive numerical dissipation \citep{Christie2024damping}.} From an atmospheric theory perspective, this jet is a result of the day--night temperature contrast driving atmospheric flow via large-scale waves. More specifically,  Kelvin waves centered near the equator travel eastward while being flanked between Rossby waves at high latitudes \citep{2011ShowmanPolvani}. In this framework, eastward momentum from high latitudes is pumped toward the equator, causing the superrotation to occur \citep{TanShowman2020}.

To first order, the emergence of this circulation regime can be understood by examining the ratio between radiative and dynamical/advective timescales \citep{PernaHeng2012, showman2013}. This ratio changes throughout a hot Jupiter's atmosphere as at deep pressures, the atmosphere is advective ($\tau_{rad}>>\tau_{adv}$), favoring jet formation. At lower pressures, the atmosphere is strongly radiative ($\tau_{rad}<<\tau_{adv}$). A short radiative timescale implies strong day--night temperature gradients \citep{2009Showman}.  While informative, several authors have noted that this timescale ratio approach leaves out important processes affecting circulation including rotation rate and drag forces \citep{PerezBecker2013,Komacek2016}. A more complex derivation and parameter exploration is outside the scope of this paper, but we refer the reader to discussions in \citealp{2011ShowmanPolvani, PerezBecker2013, Tsai2014, Komacek2016, Showman:2020rev, Debras2020, Komacek2025hysteresis, akin2025}.

Overall, the combination of strong day--night forcing and relatively slow rotation rate result in the formation of a super-rotating equatorial jet. The specific characteristics of that jet are governed by a variety of planetary processes. For example, the exoplanet's rotation rate plays a key role in jet formation and size. Very rapid rotation ($P \lesssim  1$ day ) can result in additional mid-latitude jets \citep{Showman2015} or westward offsets \citep{Zhan2024bat}, while slow rotation  ($P \gtrsim 10$ days) can weaken or completely disrupt the jet \citep{TanShowman2020, Beltz2021}. 

The influence of planetary rotation on jet size is best explained by the  \textit{Rossby Deformation Radius} \citep{Showman2002}, given by:
\begin{equation}
    L_{eq}=\bigg{(}\frac{NH}{\beta}\bigg{)}^{0.5}
\end{equation}
where $\beta = \frac{df}{dy}$ and $f$ is the Coriolis parameter, a function of latitude,  $f=2 \Omega \textnormal{ sin(}\phi)$ \footnote{Outside the equator, the deformation reduces to $\frac{NH}{f}$.}. The numerator, $NH$, is the product of the Brunt-Väisälä frequency and the atmospheric scale height. This product represents the speed of a wave traveling horizontally \citep{Showman:2020rev} and for a hot Jupiter is often more than half of the planetary radius.  The deformation radius represents the ratio between buoyancy forces and Coriolis forces \citep{PerezBecker2013} and influences jet width. A faster rotation rate/shorter orbital period will result in thinner jets. At rapid enough rotation, the higher latitude jets can form, weakening the main equatorial jet strength \citep{akin2025}.

Atmospheric temperature also plays a key role in shaping the jet. Given the tidally locked nature of these systems, higher equilibrium temperatures are correlated with faster rotation rates. For a constant stellar temperature, increasing the equilibrium temperature (and thus moving to slightly faster rotation) results in a faster jet speed due to higher eddy velocities. When the exoplanetary equilibrium temperature is constant, increasing  the stellar temperature (and thus decreasing the rotation rate) results in weaker jets \citep{Tan2024}. When planetary rotation is fixed, increasing the equilibrium temperature of the planet decreases the jet strength \citep{Tan_2019}. \citet{akin2025} find for their suite of modeled planets ranging from the cooler hot Jupiter HD~189733b to the ultrahot Jupiter WASP-121, the jet strength decreases as equilibrium temperature and rotation rate both increase.

Atmospheric metallicity will also shape the jet, as higher mean molecular weight atmospheres have longer radiative and advective timescales and result in stronger day--night temperature contrasts. At higher mean molecular weights, both the jet wind speed and width decrease \citep{Zhang2017}. \footnote{This trend may not extend to sub-Neptunes, as \cite{2018Drummond} found that for their GCM models of the sub-Neptune GJ1214~b, increasing the metallicity had the opposite effect on the jet---making it faster.}  For UHJ, GCMs that include hydrogen dissociation---which decreases the day--night temperature contrast---also have weaker jet strengths \citep{Tan_2019}. Higher-gravity GCMs show jets extending to deeper pressures \citep{Rothgrid2024}.

Finally, specific model assumptions will alter jet strengths. \cite{Komacek2022} found increasing the flux upwelling from the bottom boundary of the model (a higher $T_{int}$, referred to as ``hot start'' in \citealt{Komacek2022}) will reduce jet speeds. Including numerical dissipation \citep{showman2013,Hammond2022,Komacek2025hysteresis} or magnetic effects \citep{RauscherMenou2013,Beltz2022a,Christie2025, Blocker2026} can weaken the jet further. 

In summary, the superrotating equatorial jet predicted in hot and ultrahot Jupiter models is a result of planetary wave interactions influenced by dayside irradiation and the relatively slow rotation rate of these objects. The jet speed and width are influenced by a combination of factors including rotation rate, temperature, atmospheric metallicity, and the inclusion of physical processes such as magnetic effects, drag, and hydrogen dissociation.

\subsection{Motivation and Structure of this Work}

In this work, we aim to explore the dynamical impact of the equatorial jet on the phase-resolved transmission cross-correlation function. We alter our jet speed and vertical extent with a 3D self consistent approach by including kinematic MHD effects as well as including less physical, but still informative models where the jet speed and spatial extent is artificially altered to explore extreme cases. This theoretical groundwork will help interpret current phase-resolved Doppler shift and width observations of UHJs, allowing future work to explore the feasibility of retrieval-type analyses on these types of observations \citep[e.g.,][]{Brogi2019,Gibson2020,Gandhi2022}.

We choose to model the standard UHJ WASP-76b under four jet strengths: strong, damped, weak, and disrupted corresponding to different strengths of MHD drag. We additionally model two synthetic jet configurations and two day--night flow configurations to isolate the influences of various dynamical processes. We describe our GCM and post-processing routine in Section \ref{sec:Methods}. We present our results in Section \ref{sec:Results}. In  Section  \ref{sec:discussion}, we provide suggestions for parameterizing CCF quantities and highlight important caveats. Finally, we provide our list of recommendations for interpreting net Doppler shifts in Section \ref{sec:conclusions}.

\section{Methods} \label{sec:Methods}

\subsection{GCM} \label{sec: GCMmethods}
We use the 3D   RM-GCM \citep{2012RauscherGCM} to model the UHJ WASP-76b. RM-GCM is a fluid dynamical model that solves the primitive equations of meteorology. This set of equations, a simplified version of the Navier-Stokes equations relevant to thin, locally hydrostatic atmospheres, describes the flow of gas on a planet \citep{Showman2008,Heng2011,Tan_2019}. The GCMs in this work were first presented in   \cite{Beltz2022a}. Two of these models (referred to in this work as the ``strong jet'' and ``weak jet'' models) have previously been post-processed in \cite{Beltz2022b} and \cite{Beltz2023} to explore high-resolution emission and transmission spectral trends, respectively. We show a summary of relevant model parameters in Table \ref{tab:gcm_params}.  Each GCM was run for 1000 planetary orbits, to ensure sufficient time to reach quasi-equilibrium in temperature and winds. The post-processed spectra shown in this work are bespoke and are described in more detail in section \ref{sec: RT}.

\begin{deluxetable}{lc}
\caption{WASP-76b GCM Parameters} 
\label{tab:gcm_params}
\tablehead{ \colhead{Parameter} & \colhead{Value}} 
\startdata
         Planet radius, $R_{p}$ & $1.31 \times 10^{8}$ m \\
         Gravitational acceleration, $g$ & 6.825 m s$^{-2}$ \\
         Orbital Period & 1.81 days \\
         Substellar irradiation, $F_{\mathrm{irr}}$ & $5.14 \times 10^{6}$ W m$^{-2}$\\
         Intrinsic heat flux, $F_{\mathrm{int}}$ & 3500 W m$^{-2}$\\
         Optical absorption coefficient, $\kappa_{\mathrm{vis}}$ & $2.4 \times 10^{-2}$ cm$^{2}$ g$^{-1}$ \\
         Infrared absorption coefficient, $\kappa_{\mathrm{IR}}$ & $1 \times 10^{-2}$ cm $^{2}$ g$^{-1}$ \\
\enddata

\end{deluxetable}

GCMs are typically coupled with a radiative transfer scheme, which range in complexity. Although the RM-GCM is capable of the more complex and computationally expensive picket-fence method \citep{Parmentier2015PF, Malsky2024PF}, we choose to use models with the simpler double-gray approach. Previously, it was shown in \cite{Beltz2024}, that for transmission spectra, the choice of radiative transfer had negligible effects on transmission net Doppler shifts. We therefore choose to use these double-gray models, as this radiative transfer technique is also more straightforward for other GCMs to replicate \citep{Heng2011, RauscherMenou20212}. 

In this work, we focus on four GCMs with varying surface magnetic field strengths, which in turn sets the drag strength and influences the corresponding jet structure. The RM-GCM uses a kinematic MHD, or ``active drag,'' approach when considering the effect of a planetary magnetic field. In this framing, we assume that the planet has a dipolar magnetic field aligned with the axis of rotation. Combined with the non-ideal MHD induction equation, we can approximate the effects of Lorentz forces as a Rayleigh drag characterized by spatially dependent drag timescale  \citep{Perna2010magdrag}:
\begin{equation} \label{tdrag}
    \tau_{mag}(B,\rho,T, \phi) = \frac{4 \pi \rho \ \eta (\rho, T)}{B^{2} |\mathrm{sin}(\phi) | },
\end{equation}
where $B$ is the chosen magnetic field strength in Gauss which is assumed constant for the model domain, $T$ is the local temperature, and $\rho$ is the density for each model grid point. $\eta$ represents magnetic resistivity,
\begin{equation} \label{resistivity}
    \eta = 230 \sqrt{T} / x_{e} \textnormal{ cm$^{2}$ s$^{-1}$}.
\end{equation}
where the ionization fraction, $x_{e}$, is calculated using the Saha equation taking into account the the first ionization potential for all elements up to nickel. For numerical stability, a minimum timescale of 0.0025 of the planet's rotational period is imposed. This drag is then applied to the horizontal momentum equation in the model with a corresponding Ohmic heating term in the thermal energy equation. For more specifics, see \cite{RauscherMenou2013}. This timescale can be interpreted as the time required for east-west winds to slow to zero, assuming no other forces are present \citep{Perna2010magdrag}. A benefit to this approach is that the drag timescale can vary by over 10 orders of magnitude (ranging from tens of seconds on the hot dayside to hundreds years on the cold nightside) at a single pressure level \citep{Beltz2022a}. This notably results in asymmetrical jets, which are stronger on the nightside than the more strongly dragged dayside.

This kinematic MHD technique has previously been applied to both hot Jupiters \citep{2012RauscherGCM} and ultrahot Jupiters \citep{Beltz2022a}. This approach uses spatially varying, temperature-dependent drag, which allows for a more physical treatment of magnetic effects than a uniform drag approach \citep{Beltz2024}, without the limiting computational complexity of a full non-ideal MHD treatment \citep{Rogers_2014b, Rogers2017}. For the purposes of this work, our kinematic MHD models allow us to vary the equatorial jet's strength and spatial extent in a physically informed way. 

We explore three non-zero values for the surface magnetic field strength, resulting in four different jet strengths. Our chosen field strength values are consistent with scaling predictions based on solar system values \citep{Christensen2009,Yadav2017} which predict a wide range of values from smaller than one Gauss to hundreds of Gauss for hot Jupiters. At the time of this publication, no exoplanet magnetic fields have been directly measured, so we explore a range of strengths varying over three orders of magnitude.  Our strongest jet occurs when the magnetic field strength is set to 0~G---that is, there is no additional drag. For our smallest field strength of 0.3~G, the jet is still present throughout the atmosphere, albeit weakened. We refer to this model as the ``damped jet.'' With a surface field strength of 3~G, the jet is destroyed for the entirety of the dayside atmosphere probed by high-resolution transmission spectra \added (roughly 1 $\mu$ bar -- 1 mbar) but is still present on the nightside---this represents our ``weak jet.'' In our strongest field strength of 30~G, no jet is present on the dayside at any pressure. We refer to this model as ``disrupted jet.''

\subsection{Synthetic Models}

As additional, simplified tests of the jets' impact, we consider two further models: one with an artificially decreased jet speed, and one with an artificially enhanced jet speed. To construct these models, we begin with the strong jet (0~G magnetic field) GCM. We then fit a simple model to the zonal-mean flow:

\begin{equation}
    u = S e^{-(\phi^2)/(2\sigma_{\phi}^2) - ((p - p_c)^2)/(2\sigma_{p,\pm}^2)},
\end{equation}
where $S$ is the strength of the jet in its core in km/s, $\sigma_{\phi}$ is the jet's width in latitude, $p$ is ($\log_{10}$) pressure,  $p_c$ is the ($\log_{10}$) pressure at the center of the jet, and $\sigma_{p,\pm}$ is the jet's width in pressure. We consider an asymmetric Gaussian such that the pressures above and below the jet center are subject to different jet widths. The motivation for this approach is that the top and bottom of the jet are regulated by distinct physical processes. The bottom boundary of the jet exists at deep pressures, where the atmosphere is dominated by flux from the interior as opposed to  stellar irradiation.  In this regime, there is no day--night contrast to trigger the standing planetary-scale waves that drive the jet \citep[e.g.,][]{2011ShomanPolvani}. At the top of the atmosphere, the jet is disrupted where the radiative timescale is shorter than the wave propagation timescale \citep{2009Showman}, restricting meridional energy transport.

Our fit to the GCM jet is adequate, with residuals on the order of 15\%. We find that the fitted jet's covering fraction of the limb, defined as

\begin{equation}
    \mathcal{F}_\text{jet} = \frac{
    \displaystyle\int_{-\pi/2}^{\pi/2} e^{-\phi^2/2\sigma_\phi^2}\,d\phi
    \cdot
    \int_{p_\text{top}}^{p_\text{bot}} e^{-(p - p_c)^2/2\sigma_{p,\pm}^2}\,dp
}{
    \displaystyle\pi\,(p_\text{bot} - p_\text{top})
},
\end{equation}

is roughly $15\%$.

With our jet fit completed, we add two models to our grid: ``no jet,'' in which the jet is subtracted from the 0~G case, and ``double jet,'' in which the jet fit is added to the 0~G case, effectively increasing the jet's strength. Of course, these models do not obey the primitive equations, as the equatorial motions are no longer in balance with the rest of the atmosphere. Even so, they may provide further intuition as to how the jet structure alone impacts phase-resolved Doppler shifts.

Finally, we seek to assess whether jets can be distinguished from other dominant forms of atmospheric circulation. We therefore consider two further synthetic models that lack jets and only include \textit{day--night winds}. These flows are responsible for a substantial portion of hot Jupiter day--night heat redistribution \citep[roughly half in][]{hammond2021rotational}. They often arise in hot Jupiter GCMs when the radiative and drag timescales are short, disrupting the jet and allowing the winds to stream along the pressure gradient \citep[e.g.,][]{2011ShomanPolvani,showman2013}. 

\cite{hammond2021rotational} showed that hot Jupiters' day--night winds can be isolated with a Helmholtz decomposition of the full horizontal wind field $\mathbf{u} = (u,v)$ \citep{dutton2002ceaseless}:

\begin{equation}
    \mathbf{u} = \mathbf{u_d} +\mathbf{u_r}, 
\end{equation}

where $\mathbf{u_d}$ is the divergent (vorticity-free) flow and $\mathbf{u_r}$ is the rotational (divergence-free) flow. In this framing, the divergent component corresponds to the model's day--night flow. We extract $\mathbf{u_d}$ from our 0~G GCM using the \texttt{windspharm} \citep{dawson2016windspharm} package's ``Helmholtz'' routine. As in \cite{hammond2021rotational}, we set the spherical harmonic truncation $n=21$. We save $\mathbf{u_d}$ as our ``day--night'' model, and we double $\mathbf{u_d}$ for a ``double day--night'' model. As with the ``no jet'' and ``double jet'' models, we stress that these models are not self-consistent. They are useful inasmuch as they provide insight into the contribution of individual circulation components to the final cross-correlation function.

\subsection{Radiative Transfer Post-Processing} \label{sec: RT}
We next transform our GCM predictions (thermal and wind structures) into observables (here, transmission spectra) with ray-striking radiative transfer.

We use the framework of \cite{Kempton2012} to calculate high-resolution ($R =250000$) optical (380~nm to 950~nm) transmission spectra. We begin by regridding the GCM from a pressure grid (originally 88~bar--10~$\mu$bar) to an altitude grid, so that the lines of sight that we later track move straight through physical space. Any cell with a pressure lower than the lowest considered (1~$\mu$bar) is omitted from the optical depth calculation. To extend the model domain to lower pressures (1~$\mu$bar), we perform isothermal extensions to the GCMs' temperature fields and barotropic extensions to the GCMs' velocity fields. We perform these extensions (i.e., copy $T$, $u$ (east-west/zonal winds), and $v$ (north-south/meridional winds) upward to lower pressures) to resolve strong line cores in our simulated spectra. Our isothermal extensions are appropriate for this study --- double-gray radiative transfer converges to an isothermal profile in the upper atmosphere \citep{Guillot2010}. Furthermore, transmission spectra are not particularly sensitive to temperature gradients. We null the vertical wind profile, as vertical velocities do not feature into our Doppler shift calculation (Equation~\ref{eq: los}).  Our barotropic extension is less grounded, as the wind speeds should in principle continue to increase due to greater gravity wave amplitudes (and thus driving) at lower pressures. We find that our results are not very sensitive to the exact choice of velocities within this extension regime (Appendix~\ref{appendix:barotropic_test}).

With our inputs organized, we next track the wavelength-dependent optical depth, $ \tau(\lambda)$, accumulated through latitude--altitude lines of sight:

\begin{equation}
    \tau(\lambda) = \int \kappa(\lambda) ds,
\end{equation}
where $\kappa(\lambda)$ is the wavelength-dependent absorption coefficient (summed over all species) and $ds$ is the differential path length. We calculate $\kappa(\lambda)$ locally assuming chemical equilibrium under solar composition \citep{Lodders2003, Fastchem2018}; each gas species’ absorption is calculated via opacity sampling \citep{malik2019self} the associated line lists of Na, K, Fe, Mg, V, and Mn \citep{kurucz1995kurucz}. We opt to not calculate opacity of our strong lines (e.g., Na, K) far out into the wings because our opacity sources neglect pressure broadening. This approximation is reasonable, as the cross-correlation function is primarily sensitive to the position of strong spectral lines.

The line lists are Doppler-shifted to account for local motions relative to the observed line-of-sight, $v_{\rm LOS}$:

\begin{equation}
\label{eq: los}
\begin{split}
     v_{\rm LOS} = -u\sin(\theta) - v\cos(\theta) 
    \sin(\phi) \\- \Omega(R_{\rm P} + z)\sin(\theta)\cos(\phi),
\end{split}
\end{equation}
for eastward velocity $u$, north--south velocity $v$, longitude $\theta$, and altitude $z$

We produce phase-resolved spectra over the course of transit. As in \cite{2022Savel}, to account for limb darkening, we alter the amount of light through the atmosphere (each cell) and decrease the opaque inner atmospheric boundary size with \texttt{batman} \citep{Kreidberg2015} calculations. We calculate spectra at 31 phases (rotating the GCM output at each phase) to resolve the change in the net \textit{projected} Doppler shift as the transit progresses. In total, the planet subtends 30.8 degrees in phase during its orbit; between each snapshot, it rotates roughly 1 degree.

\subsection{Contribution Functions} \label{sec: contribution}
Interpreting our spectra relies on understanding which pressure levels they probe. To this end, we calculate spectral contribution functions. 

Defining the contribution function in 3 dimensions is nontrivial. To simplify the calculation and its interpretation, we therefore perform this assessment in 1D. We do so by first producing a 1D average of the 3D thermal structure, by averaging the GCM output over all latitudes. We collapse the longitudinal dimension by averaging over a select set of longitudes that most impact the transmission spectrum---the ``opening angle.'' Using the formalism of \cite{Wardenier2022}, the opening angle of WASP-76~b is  $\approx 24^{\circ}$, centered on the terminator.

We then post-process this averaged thermal structure with \texttt{the petitRADTRANS} \citep{molliere2019petitradtrans} 1D transmission spectrum code, using the same chemistry as our 3D radiative transfer code. We assume no spectral line broadening due to winds or rotation.

With our nominal transmission spectra calculated, we calculate the contribution function at each atmospheric layer, $C^i_{\rm tr}(\lambda)$, following Eq. 8 from \cite{molliere2019petitradtrans}:

\begin{equation}
    C^i_{\rm tr}(\lambda) = \frac{\Delta_{\rm nom} (\lambda)- \Delta(\kappa_i(\lambda)=0)}{\Sigma_{j=1}^{N_L}[\Delta_{\rm nom}(\lambda) - \Delta(\kappa_j(\lambda) = 0)]},
\end{equation}
where $\Delta_{\rm nom}$  is the nominal transmission spectrum and  $\Delta(\kappa_i(\lambda)=0)$ is the transmission spectrum calculated when all the opacity in only the $i$th atmospheric layer is set to 0.  We repeat this step for each of the $N_L$ layers; $N_L=90$ for these spectral calculations. In sum, the numerator in this equation describes how the $i$th layer contributes to the transmission spectrum, and the denominator normalizes over contributions from all layers.

\begin{figure*}
    \centering
    \includegraphics[width=0.95\linewidth]{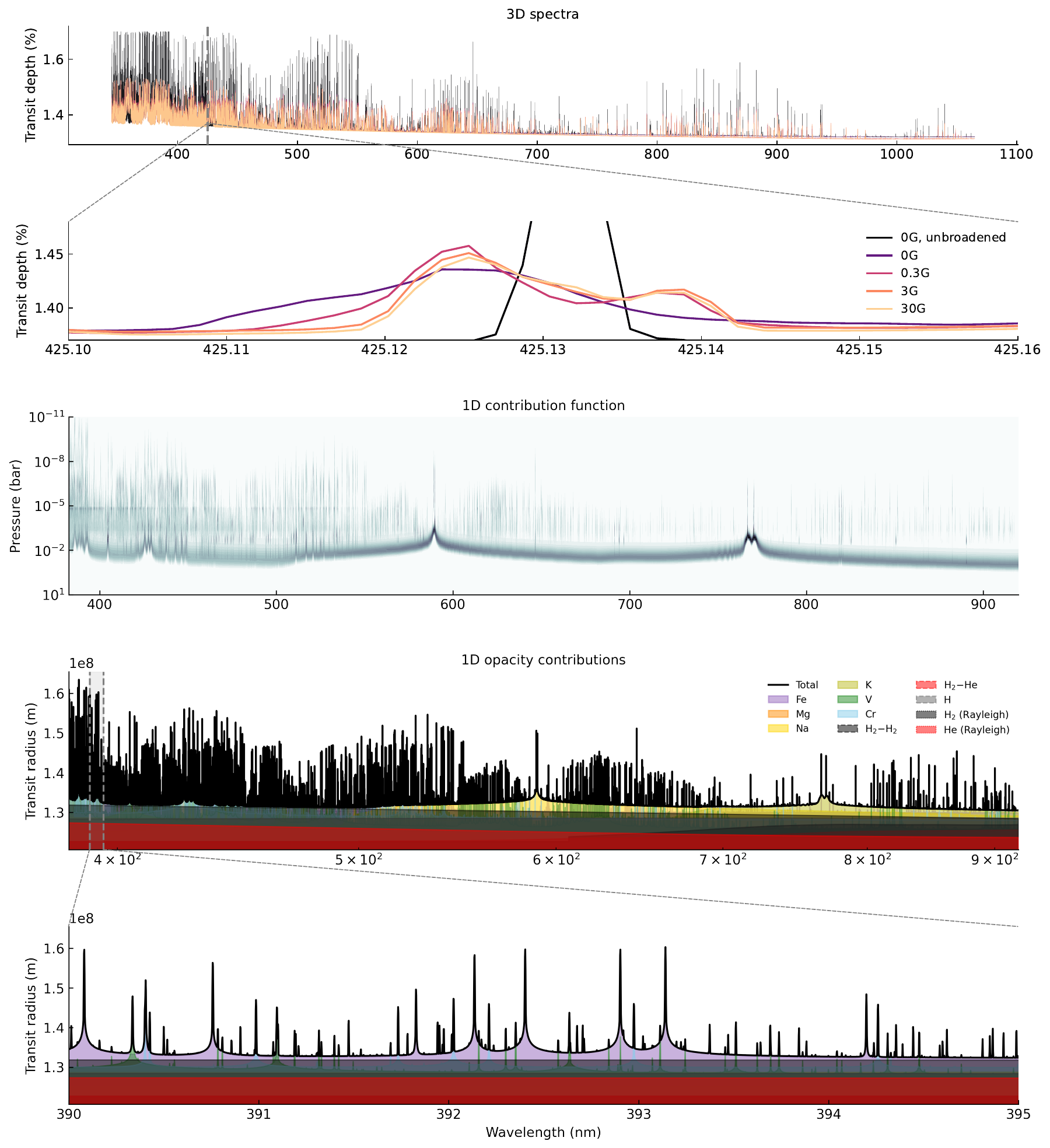}
    \caption{Several representations of the spectra used in this work. Top panel: the unbroadened and broadened spectra as calculated with the 3D radiative transfer code. Second panel: the same spectra, zoomed in on a single spectral line to reveal the impact of different magnetic field strengths. Note that both blue shifted and redshifted components arising from the 3D model are visible. Third panel: the 1D contribution function calculated with \texttt{petitRADTRANS}. Fourth panel: the 1D opacity function, indicating which species dominates at each wavelength. Fifth panel: the same opacity function, zoomed in on a narrow spectral region dominated by iron lines.}
    \label{fig:contribution}
\end{figure*}

These calculations indicate that our spectral \textit{lines} generally probe pressures between 0.1~mbar and 1~mbar, with the strongest line cores nearing 0.1~nanobar (Fig.~\ref{fig:contribution}). Note that in cross-correlation, we are not sensitive to the continuum opacity that may probe slightly deeper pressures---only the strongly wavelength-dependent spectral lines.

\subsection{Cross-correlation}

We next calculate the cross-correlation functions associated with these spectra with the \texttt{scope} package \citep{Savel2025}.\footnote{\url{https://github.com/arjunsavel/scope}} We cross-correlate the model with a \textit{template spectrum} that has no velocity field associated with it --- that is, the center of transit spectrum with Doppler effects turned off. These template spectra include opacity only from Fe I to better match the single-species tests performed on real data. The velocity step size of this calculation is 0.5~km\,s$^{-1}$. Because we are not comparing our simulations to any specific observation, we do not include any noise or model distortions due to data cleaning.

Once we calculate our cross-correlation function for each spectrum, we determine the net planet-frame velocity of the spectrum with respect to the template. We do so by fitting a Gaussian function to the cross-correlation function. We take the mean of this Gaussian as the net velocity, and we take the variance as the broadening. This approach is typical for phase-resolved measurements of Doppler shifts in transmission spectroscopy \citep[e.g.,][]{simonnin2025time}.

\section{Results} \label{sec:Results}
\subsection{GCM Results}
\begin{figure*}
\centering
\includegraphics[width=.80\linewidth]{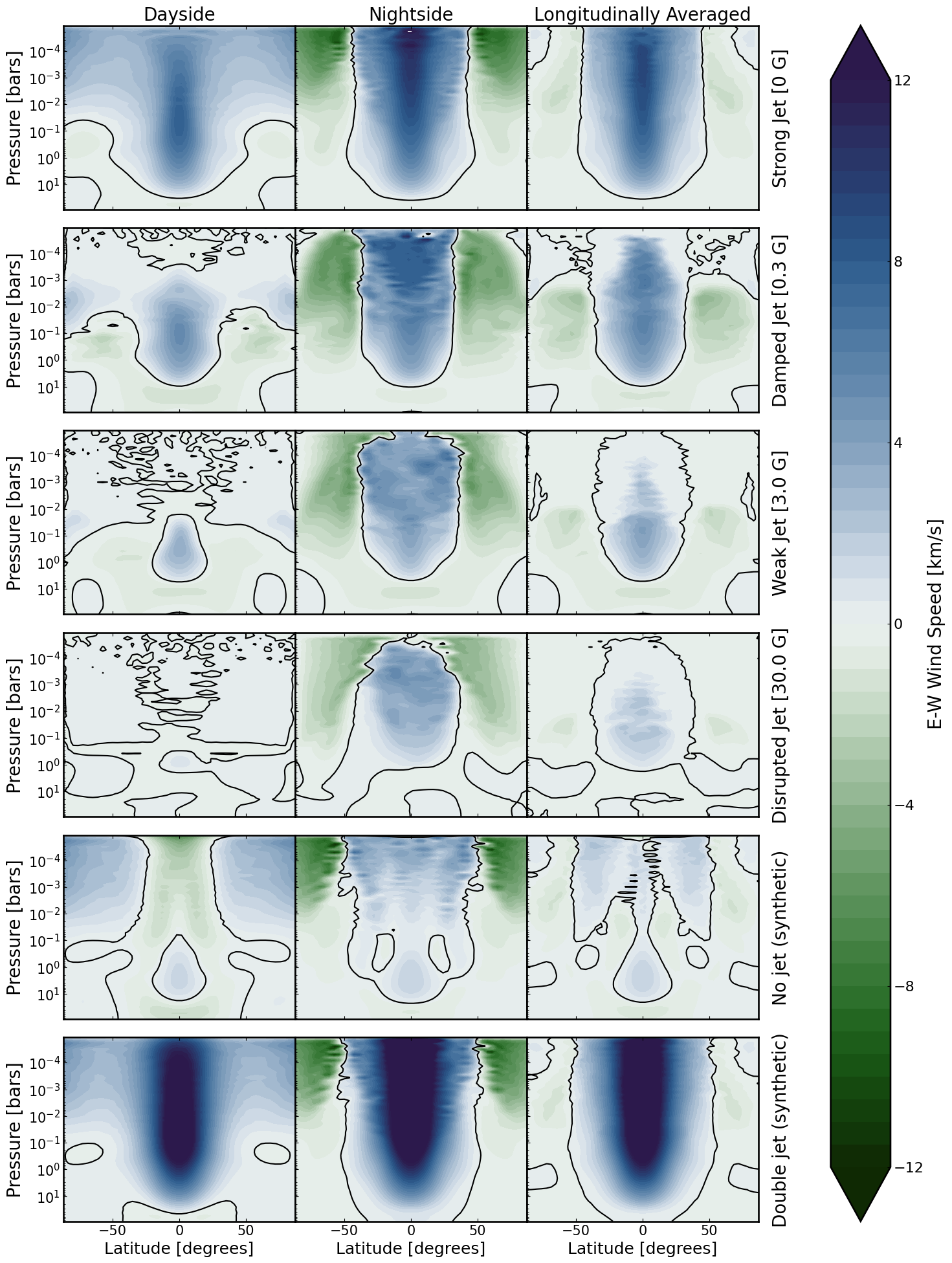}
\caption{Our four different GCM models have varying strengths of jets due to the different levels of assumed $B$ field strength. Here, we show dayside, nightside and  longitudinally averaged zonal (eastward) winds from the four GCM models explored in this work. In the top row, the model without any active drag has the strongest jet, which is present for most of the atmosphere. As the magnetic field strength in our active drag routine is increased, the jet becomes slower and spans a smaller range of pressures. For our active drag models in the subsequent three rows, it should be noted that the jets, regardless of strength, are not symmetric with longitude due to the temperature dependence of the kinematic MHD approach. The jets are stronger on the nightside of the planet, where the temperatures are lower and thus the drag is weaker \citep{Beltz2022a}. The bottom two rows show the zonal-mean zonal wind of our synthetic jet models, which subtract and double the strength of a jet profile fitted to the 0~G model, respectively.}
\label{fig:jetsallmodels}
\end{figure*}

In Figure \ref{fig:jetsallmodels}, we show our four self-consistent models of interest and their respective zonal flow on the dayside, nightside, and averaged over all longitudes. Our strong jet case, which has no additional drag beyond small numerical dissipation required for simulation stability, shows a strong jet that reaches over 12~km\,s$^{-1}$ on the nightside and nearly 8~km\,s$^{-1}$ on the dayside extending throughout nearly the entire modeled atmosphere. Our three cases that include kinematic MHD show various degrees of jet disruption. Due to the temperature-dependent nature of the active drag, these jets are more strongly dragged on the dayside than the nightside. At a small (0.3 G) B field strength, the model is still able to retain a weak jet, reaching 11~km\,s$^{-1}$ and  5~km\,s$^{-1}$ on the nightside and dayside respectively. However, the jet is not maintained at the lowest pressures modeled, as the drag timescale (Eq.~\ref{tdrag}) decreases linearly with density---which decreases exponentially toward the top of the atmosphere following hydrostatic equilibrium. The trend continues for the more  strongly dragged weak jet case at 3~G,  in which the jet is only substantially present in the deep atmosphere on the dayside, deeper than  is probed by our transmission spectroscopy (Fig.~\ref{fig:contribution}). Here, the dayside jet only reaches speeds of 3.5~km\,s$^{-1}$, and the nightside reaches speeds roughly double that.  Finally, at 30~G, the Lorentz force is strong enough throughout the entire atmosphere such that no jet structures are able to form on the dayside, as these winds do not even reach 1~km\,s$^{-1}$. A very weakened jet is able to be maintained on the nightside, with speeds reaching up to  5.5 ~km\,s$^{-1}$, but at deeper pressures than probed with our model spectra.    

\begin{figure*}
    \includegraphics[width=7 in]{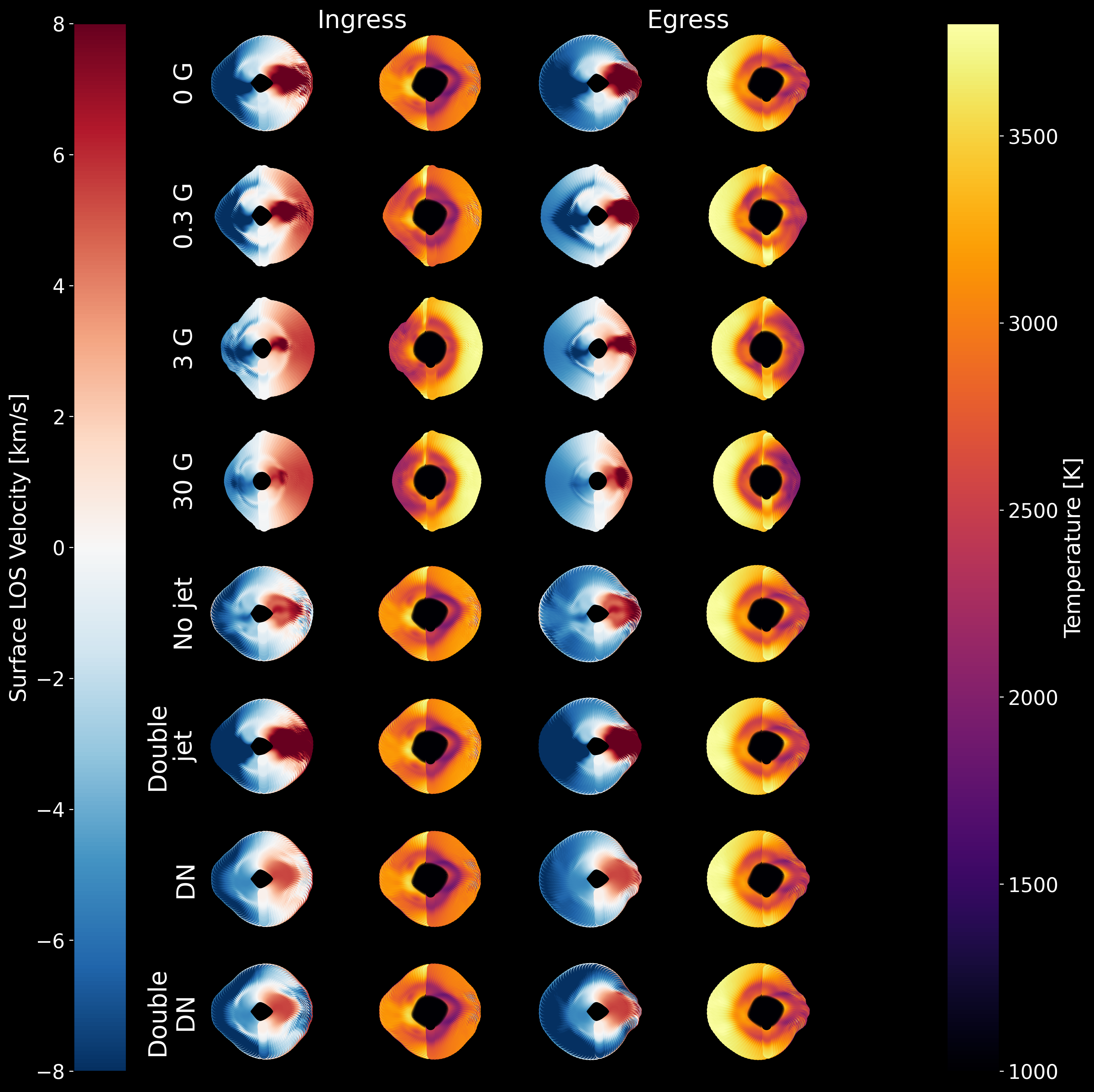}
    \caption{Differing drag strength results in significant changes in the limb temperature and jet structure. Here we show the line of sight velocity---including both winds and rotation---as well as temperature structure of the east and west limbs (the left and right half of each annulus respectively) from our four models at ingress and egress. We additionally show the velocity structure from our models with the jet artificially removed and enhanced. Finally, the bottom two rows show the synthetic models with purely divergent flow (with day--night abbreviated as DN). Due to the short orbital period of the planet, there is significant rotation during transit, resulting in the limbs varying strongly from ingress to egress. }
    \label{fig: modellimbs}
\end{figure*}

To visualize model differences during transit, we show line-of-sight velocity (including both winds and rotation) and temperature projections for both limbs of each model during ingress (from $-15^{\circ}$-- $-11^{\circ}$) and egress (from $11^{\circ}$-- $15^{\circ}$) in Figure~\ref{fig: modellimbs}.  Notably, these models are plotted in altitude space, so the extent of the atmosphere is determined by variations in the atmospheric scale height. These variations are in turn driven by changes in the local temperature \citep[e.g.,][]{brown2001transmission}. The kinematic MHD models have stronger day--night temperature contrasts, which results in stronger differences in the spatial extent of the modeled atmosphere. The strong jet of the 0~G model is apparent during both ingress and egress. During ingress, there is a ``ring'' of near-zero line of sight velocity in all models. However, the mechanisms causing this ring differ. In the 0~G case, this ring is a~ result of high-latitude gyres providing net momentum in the opposite direction of the jet. In the 3~G case, these nightside gyres are unable to form. Instead, this is a result of the poleward flows caused by the active magnetic drag routine. This is also true for the 30~G case, though the ``ring'' itself is harder to see due to the overall lower line-of-sight velocities. The 0.3~G case shows evidence of each of these behaviors. Small gyres are able to form, but they are much weaker than the ones present in the 0~G case and are aided by the poleward flow to generate the same ``ring'' structure.

\subsection{Cross-Correlation Results}
\subsubsection{Kinematic MHD Models}

In Figure~\ref{fig:ccfs}, we show the phase-resolved Doppler shifts from our 0~G, 0.3~G, 3~G, and 30~G models as a function of phase in the top panel. For all models, as transit progresses, the net Doppler shift becomes more blueshifted as the hotter eastern limb rotates into view. Due to the damping of winds in the active drag models, the  maximum  and minimum line of sight velocity are smaller in magnitude than the drag-free case. Interestingly, the 0.3~G model is very similar to the 0~G model, implying that this level of magnetic field strength may be too weak to significantly alter the circulation  of the planet, at least at the limbs. The 3~G and 30~G models are very similar and notably switch from redshift to blueshift much later in phase compared to the 0~G model. Thus, strong drag causes the crossing from red to blueshifted velocities to occur much closer to mid-transit. This is due to the asymmetric, weaker jets taking longer to impart a net blueshift compared to our strong jet models with the same planetary rotation.

\subsubsection{Synthetic Jet Models}
\begin{figure}
    \centering
    \includegraphics[width=0.99\linewidth]{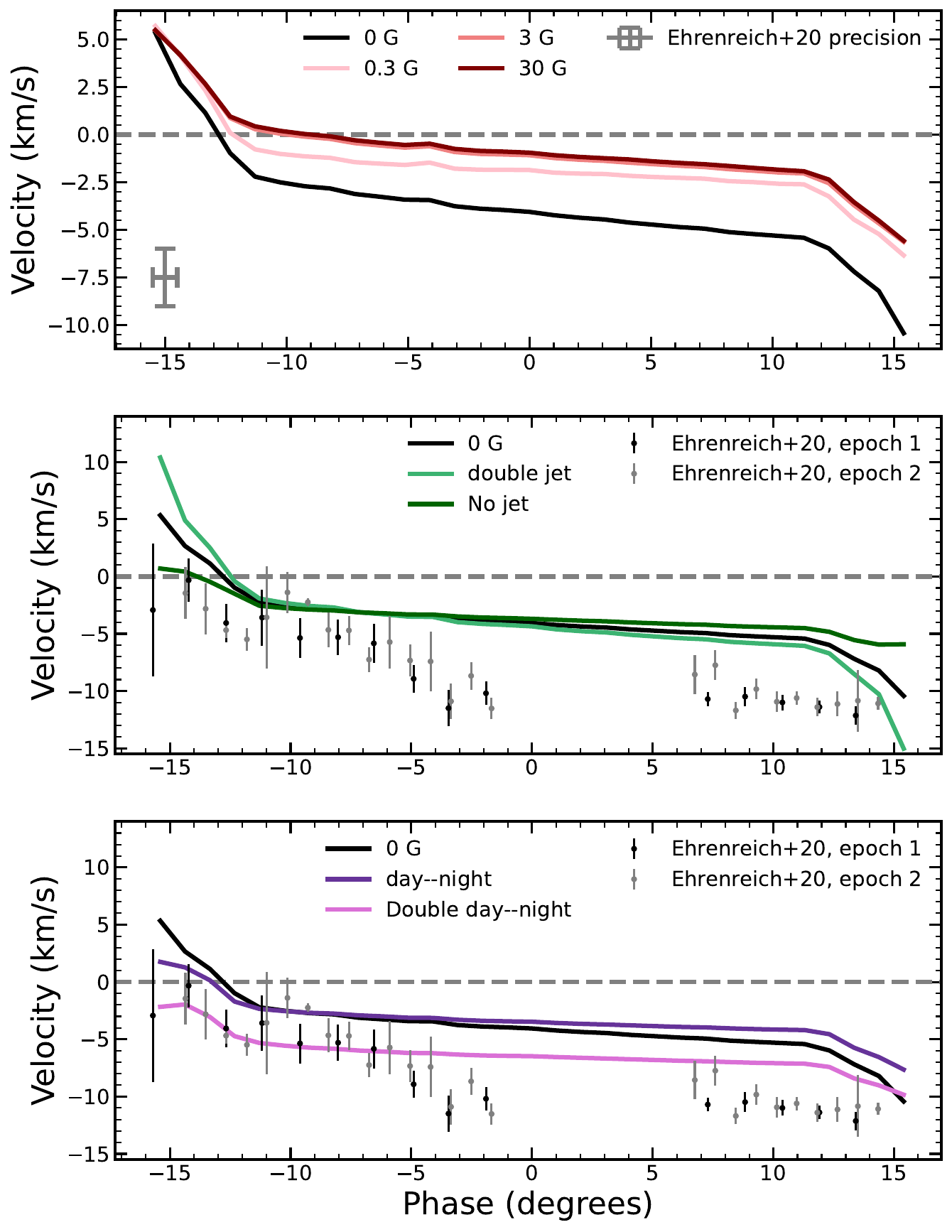}
    \caption{Phase-resolved Doppler shifts from our post-processed GCMs as a function over the course of transit. The top panel shows these results for our GCMs, the middle panel shows these results for the synthetic jet models, and the bottom panel shows these results for the synthetic day--night wind models. The inclusion of magnetic drag weakens the magnitude of the retrieved net Doppler shifts and results in a later crossing from red to blueshift compared to the drag-free model. The synthetic jet models demonstrate that the strongest influence of jet strength occurs during early ingress and late egress. The top panel shows the approximate precision of a single transit from \cite{Ehrenreich2020} to demonstrate the typical data quality of high-SNR phase-resolved spectra; the middle and bottom panels show those data in full. Note the variation in y-axis scale between the panels.}
    \label{fig:ccfs}
\end{figure}

We show the effect of our synthetic jet configurations in the middle panel of Figure \ref{fig:ccfs}. As expected, increasing the jet strength (the ``double jet'' scenario) results in stronger net Doppler shifts. At this jet strength, which reaches speeds up to 22~km\,s$^{-1}$---much faster than any published GCM---the net Doppler shifts in egress reach larger values than reported in \citet{Ehrenreich2020}. However, our ingress differs strongly from \cite{Ehrenreich2020}, who inferred a net blueshift at all phases. This suggests that the mismatch between wind speeds in GCMs and high observed net Doppler shifts is likely more complex than over-damping in hot Jupiter GCMs. Removing the jet results in a very small change in net Doppler shift throughout orbit, almost entirely due to rotation. 

For a majority of transit (roughly $-10\degr$ to 5$\degr$), however, the models show similar net Doppler shifts. The strength of the jet therefore is most influential during the beginning of ingress and end of egress. This behavior is due to the fact that during these phases, the planetary limbs scan across the stellar disk, ``spatially resolving'' the terminator \citep[e.g.,][]{Kempton2012}. The net Doppler shifts at these phases are therefore averaged over fewer lines of sight, allowing the large line-of-sight velocities near the edge of the planetary disk to dominate the signals at the beginning of ingress and the end of egress.

\subsubsection{Synthetic Day--night Models}

Unlike the superrotating jet, the day--night models do not impart a slope to our models' Doppler shifts during transit. Rather, the day--night wind adds an offset to the Doppler shifts caused by planetary rotation and thermal asymmetry. This effect can be understood as follows: The projected jet speed is strongly asymmetric, so a thermal asymmetry that increases with phase will contribute an increasing amount of one jet component into the net Doppler shift as a function of phase. In contrast, the day--night flow is much more symmetric across the limbs, so the fractional contribution of each limb to the net Doppler shift does not greatly change the net Doppler shift.

The day--night only model's Doppler shifts are quite similar to our no-jet synthetic model's. The two do not exactly overlap, however, with some slight differences at ingress and egress. This departure is expected, because the jet is not the only rotational component; standing waves (e.g., Rossby gyres at the mid-latitudes) are also a substantial portion of hot Jupiters' rotational circulation component \citep{hammond2021rotational}.

\subsubsection{Removing the Impact of Rotation}
\begin{figure}
    \centering
    \includegraphics[width=0.95\linewidth]{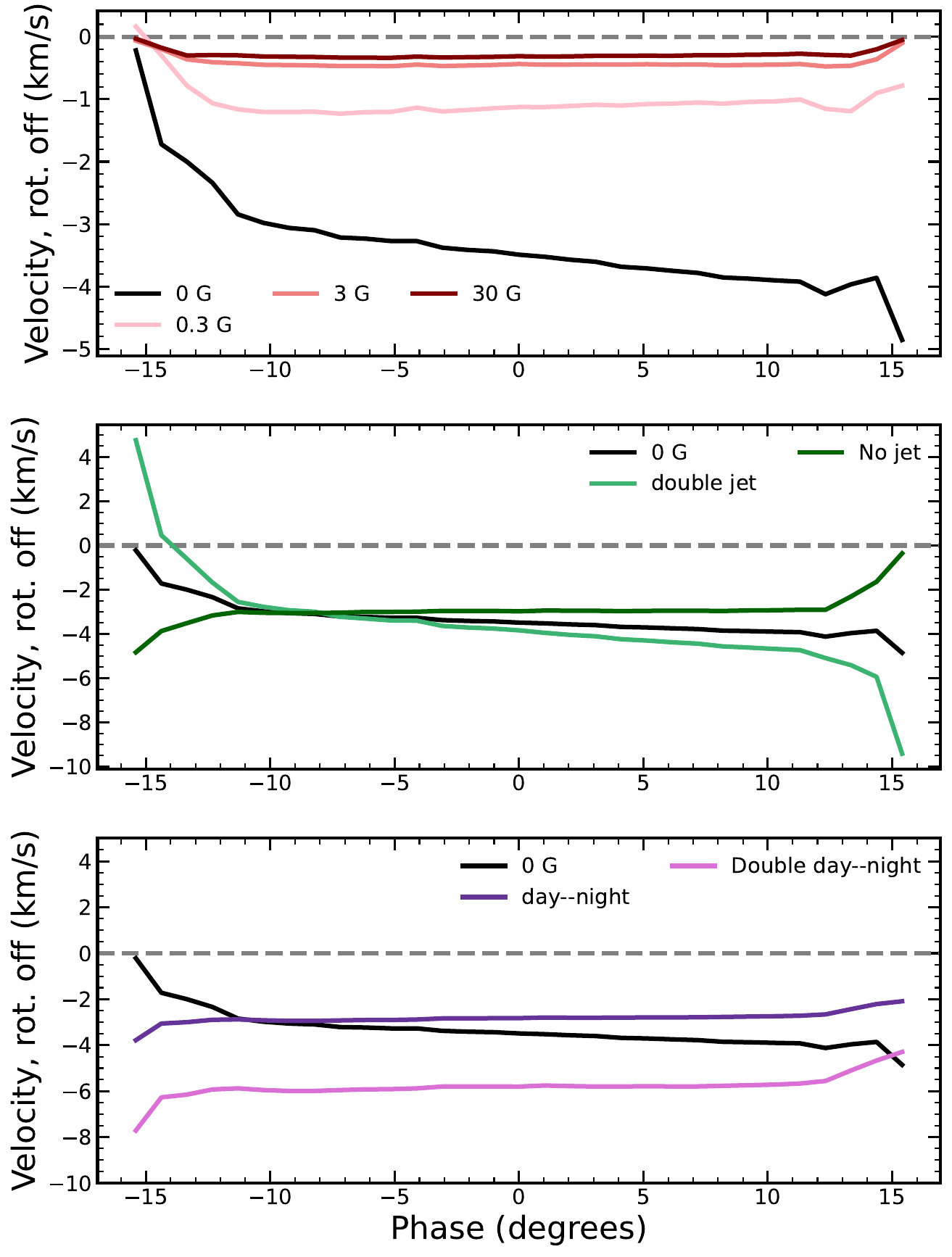}
    \caption{Same as Fig.~\ref{fig:ccfs}, but removing the effect of planetary rotation in our broadened spectra. Without rotation, our strongly dragged models with the weak and non-existent jets have very small shifts during the entire transit.  }
    \label{fig:ccfs_norot}
\end{figure}

Up until now, each post-processed spectrum has included the effects of both winds and rotation in its line broadening (Eq.~\ref{eq: los}). The amount of broadening caused by rotation is similar for all models, as they all share the same rotation rate.\footnote{Notably, the broadening is not exactly the same, due to the presence of $z$ in equation \ref{eq: los}. The hotter models have larger scale heights, resulting in different altitudes at the same pressure level.} To isolate the effects of winds, we can omit the rotational broadening term in Equation~\ref{eq: los} to isolate the impact of planetary winds, which is shown in Figure \ref{fig:ccfs_norot}. Even without accounting for rotation, we see that the 0~G model shows a similar trend as we have seen, becoming more blueshifted as transit progresses. The weakest jet cases of 3 and 30~G, however, show a roughly constant small \textit{blueshift} for nearly the entire transit. This is due to their magnetically dominated circulation, which results in a greater portion of each limb being redshifted during transit (see Figure \ref{fig: modellimbs}).   Similar to Fig.~\ref{fig:ccfs}, the 0.3~G case's behavior lies between the 0~G case and strongly dragged models.

We also explore the effect of removing rotation from our synthetic jet models, as shown in Figure \ref{fig:ccfs_norot}. This allows us to measure the corresponding velocity originating from a perfectly symmetric jet; These are actual differences in flow in ingress and egress as well as tracing geometric projections  during mid-transit. A useful comparison to make is that of the ``no jet'' model in this figure and the 3~G and 30~G models in Figure~\ref{fig:ccfs_norot}. These three models have a lack of jet in common. However, when rotation is removed, the ``No jet'' model is significantly blueshifted, while the 3~G/30~G models are only very slightly blueshifted. The inability of this approach to remove the longitudinally asymmetric jet in our kinematic MHD models (Fig.~\ref{fig:jetsallmodels}) shows that the kinematic MHD approach shapes the circulation in a more complex way that cannot be approximated with a simple jet subtraction.  

Removing rotation also makes the offset component of the day--night flow evident. These models have essentially flat Doppler shifts as a function of phase, with slight deviations due to varying strength of the day--night flow. These spatial variations are made most evident at ingress and egress, which spatially resolve the limbs.

The results of these two tests confirm the strong influence of jet strength on ingress/egress. We also note that, in the case of no rotation,  the strength of the jet directly influences the slope of the net Doppler shifts during mid-transit. That is, in our modeling suite, strong jets still show a blueward slope, while models that exhibit weaker jets show a near-flat trend. 

\begin{figure*}
    \centering
    \includegraphics[width=0.95\linewidth]{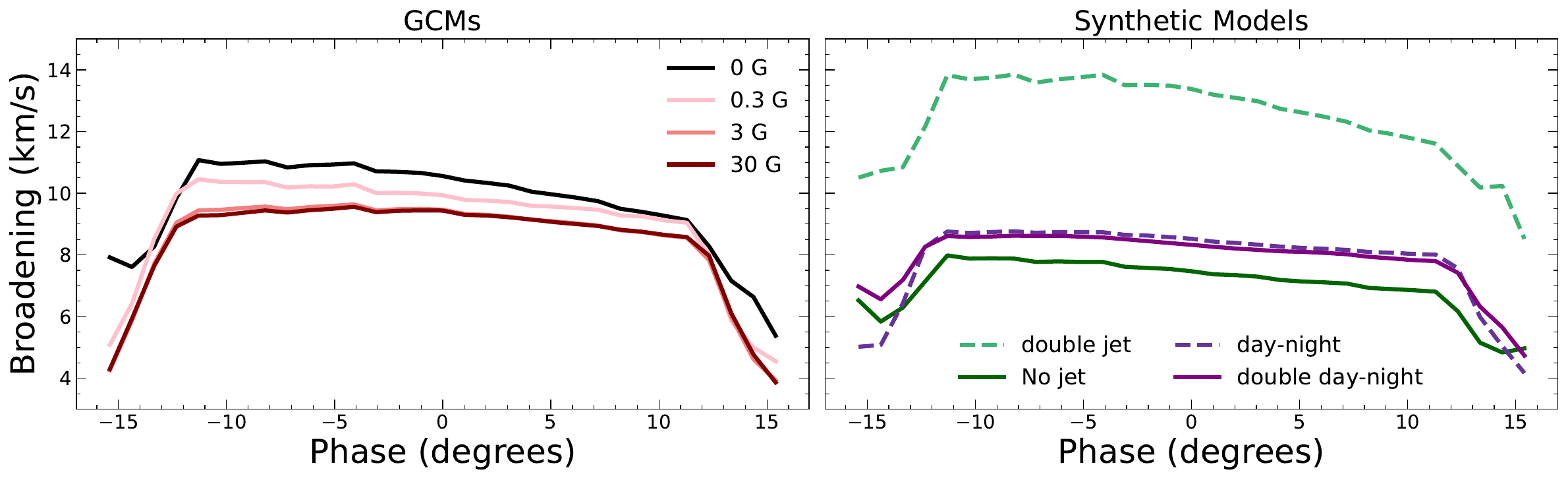}
    \caption{The broadening of the CCFs for each of our models as a function of phase. The left side shows these results for our post-processed GCMs; the right shows the results for our synthetic models.}
    \label{fig:sigmas}
\end{figure*}

\subsubsection{CCF broadening}
Given that the superrotating jet may impact the width of the CCF, we also compare the broadening of our models, both synthetic and self-consistent. We show the standard deviations of the Gaussians fit to our CCFs in Fig.~\ref{fig:sigmas}.

Our synthetic models indicate the behavior of limiting-case CCF broadening. The double-jet model exhibits the most broadening, as it has the maximal dispersion in projected velocity between the limbs. In contrast, the no-jet model exhibits the least, as the net velocity over the terminator is more homogeneous. The magnitude of the day--night flow does not impact the broadening, as shown by the similarity of the day--night and double day--night models' broadening; altering the strength of symmetric flow maintains the same amount of velocity dispersion across the limb.

Also of note is that the increased velocity dispersion in the double-jet case leads to an increased \textit{slope} in the broadening as a function of phase. We can make sense of this trend with the semi-analytic exploration of CCF width in Appendix~\ref{appendix:derivation}. There, we show that for a thermal structure that varies as a sinusoid longitudinally, increasing the velocity dispersion by a constant factor will increase the slope of the Doppler shift. We see the same effect here: The double-jet and no-jet model share the same thermal structure, but the increased difference in net velocities between the double-jet models' limbs makes a slope in the Doppler shift. The absolute CCF width is set by the velocity dispersion (among other broadening parameters); the CCF width's slope traces inhomogeneities in thermal and/or velocity structure over the angle of rotation during transit.

This intuition helps us interpret the GCMs in the left panel of Fig.~\ref{fig:sigmas}. The 0~G model has the largest thermal inhomogeneities and the strongest jet, so it has the largest CCF broadening and greatest slope in that broadening as a function of phase. The 30~G model, with the smallest inhomogeneities and most disrupted jet, has a lower broadening velocity that is relatively uniform in phase. 

Finally, all models show a drop-off in CCF width in ingress and egress. This behavior is explained by the fact that during these phases, a smaller portion of the exoplanetary limb is being sampled with a more coherent velocity distribution, and the CCF is more narrowly defined around these values.

\section{Discussion} \label{sec:discussion}
\subsection{Parameterizing CCF quantities}\label{sec:synthetic}

Our synthetic jet models confirm that the jet strength impacts Doppler shifts in two ways: ingress / egress variations and a slope during mid-transit. The ingress and egress variations clearly contain the most information about the jet; this behavior echoes the findings of space-based limb asymmetry work, which finds that ingress and egress phases contain the most information about limb asymmetries \citep[e.g.,][]{von2016inferring,kempton2017observational, powell2019transit}. These changes are faster than linear and may not be simple to parameterize in HRCCS retrievals \citep[e.g.,][]{Brogi2019, Gibson2020}. Furthermore, the ingress and egress phases are necessarily at lower signal-to-noise than exposures during the full transit, as less of the planet's atmosphere is occulting the star. To first order, this decrease in SNR follows the ratio of the phase-dependent occultation to the full transit depth. By symmetry, the \textit{average} ingress SNR is half the SNR of a full-transit exposure, with the SNR increasing as more of the planet occults the star. This intuition is approximately validated in the \cite{Ehrenreich2020} phase-resolved observations (Fig.~\ref{fig:ccfs}). The intrinsic difficulty with this approach is as the SNR increases, the spectrum averages over more lines of sight, bringing the net measured Doppler shift closer to the limb average. That is, while the CCF SNR increases as ingress progresses, the ``inhomogeneity signal'' decreases.

There may be more hope for the mid-transit phases’ Doppler shifts. When the full planetary disk is occulting the star, our synthetic jet models show that the slope of the Doppler shifts varies linearly with the (synthetic) jet speed. This effect compares favorably to observational precision. For instance, if the phase-resolved CCFs from four transits at the \cite{Ehrenreich2020} precision were constructively stacked, simple MCMC experiments indicate that the precision of the inferred mid-transit slope (0.03~km\,s$^{-1}$\,degree$^{-1}$) would substantially exceed the spread between the synthetic models that we present (0.12~km\,s$^{-1}$\,degree$^{-1}$). 

This sloping effect is distinct from the day—night models’ behavior. Because the day—night flow is symmetric across the limbs, it adds a roughly constant offset to the Doppler shifts. Small changes in the net Doppler shifts are expected due to spatial variation in this day—night flow (e.g., as isolated during ingress and egress), but this change is expected to be weaker than the change induced by a jet.

With this behavior noted, and assuming that scale height differences drive this slope \citep{Wardenier2021}, we can derive a heuristic expression for the slope of the Doppler shifts during transit (see Appendix~\ref{appendix:derivation}):
 
\begin{equation}
    \begin{split}
        \frac{d(v_{net})}{d\phi} = \bigg{(}\frac{2v_{rot} + v_m + v_e}{2T_0}\bigg{)}\cos(\theta_{offset} + \phi)\Delta T.
    \end{split}
\end{equation}

 for a sinusoidal longitudinal thermal structure with offset $T_0$, day--night contrast $\Delta T$, and offset from the substellar point $\theta_{offset}$ at some phase $\varphi$. In this framing, the rotational velocity $v_{rot}$, planet-frame morning velocity $v_m$, and planet-frame evening velocity $v_e$ are all spatially constant. We note that this is a first-order approximation, and the \cite{Ehrenreich2020} data illustrate departures from linearity; however, other phase-resolved datasets appear more linear \citep{Borsa2021, simonnin2025time}. The broad takeaway is that given a thermal inhomogeneity, velocity differences increase the magnitude of the slope. Conversely, in the absence of a thermal asymmetry, even strong winds or rotation produce little phase-dependent slope, so long as the winds do not change over the course of the observation. This is only a first-order approximation, but it may serve as a useful interpretational framework for the evolution of Doppler shifts.

In this framing, thermal and velocity gradients degenerately affect the Doppler shifts’ slope. This degeneracy may be lifted by considering the CCF width as well. In Appendix~\ref{appendix:derivation}, we show that the magnitude and slope of the CCF width are sensitive to velocity and thermal differences. This intuition also bears fruit in our models, which show substantial differences in CCF width based on synthetic jet speed and drag strength. Both the semi-analytical and numerical approaches indicate a roughly linear change in CCF width as a function of phase. Between these two observables, the two quantities of interest (thermal and velocity gradients) may be disentangled.

To leading order, we expect that the impact of heterogeneous thermal and velocity structures will not manifest strongly in variations of the CCF amplitude (Appendix~{\ref{appendix:derivation}}). However, our approach ignores questions of model--data mismatch and sharp gradients in mean molecular weight, which are both particularly relevant for ultra-hot Jupiters \citep[e.g.,][]{parmentier2018}.

In sum, the slopes and offsets of the CCF width and shift contain distinct information about thermal and dynamical inhomogeneities. Parameterizing these quantities as linearly changing observables may prove fruitful in terms of diagnosing the nature of jets with HRCCS.

\subsection{Caveats}\label{sec:caveats}

 None of the models shown here are able to reproduce the magnitude of the egress Doppler shifts from \cite{Ehrenreich2020}---which reach over 12~km\,s$^{-1}$---except for our artificially enhanced double jet model.  Currently, 3D GCMs struggle to produce the highest-magnitude Doppler shifts found in data. This is likely due in part to these  models requiring some artificial damping to remain numerically stable. Models run at higher spatial resolutions may alleviate this \citep{Heng2011} and should be explored in future work.

Near-infrared spectroscopy has also demonstrated great sensitivity to atmospheric dynamics in hot Jupiters \citep[e.g.,][]{nortmann2024crires,Wardenier2024}. While our optical spectra do not extend to this spectral region, our general takeaways should apply to near-infrared spectroscopy---provided there are enough spectral lines of varied depths to appropriately sample the jet's velocity field. Two exceptions are species that may experience strong chemical transitions in the atmosphere (e.g., $\rm H_2O$ in an ultra-hot Jupiter) or that probe very low pressures (such as the Na line core).

Although we have chosen to model only one planet, we argue that these interpretations (in particular the semi-analytic ones) should apply to nearly the entire class of UHJs. Notably, these planets nearly all have similar rotation periods --- roughly 2 days or so, putting them in the rapid rotating regime. It is not until rotation rates reach extremely short periods (that is, roughly 0.5 days) that atmospheres fall into the ``bat rotator'' regime characterized by a westward, subrotating jet \citep{Zhan2024batrotator}. There are no known ``typical'' UHJs that fall into this category, but the recently characterized pulsar planet appears to fall in this regime \citep{ZhangPulsar}. Our conclusions do not extend into the warm Jupiter regime with periods exceeding 10 days,  as the decreased day--night temperature gradients lead to more spherically symmetric atmospheres \citep{kempton2014high}.

We also note that we have assumed tidal locking for our models. Non-synchronous rotation can still lead to the formation of equatorial jets in UHJs, provided that the rotation rate remains in the same rapid rotator regime \citep{RauscherKempton2014}. A degeneracy arises here as a slightly slower rotation rate can lead to faster wind speeds resulting in similar levels of overall Doppler broadening \citep{kempton2014high,Flowers2019, Beltz2021}. 

We have decided to not include clouds in our models for the purpose of this work. Clouds are indeed expected to form on the nightsides of UHJs \citep{parmentier2021cloudy,Roman_2021}, and earlier work demonstrated that clouds could impact the shape of phase-resolved transmission spectra \citep{2022Savel}. However, HRCCS is generally less sensitive to clouds than space-based spectroscopy \citep{kempton2014high, gandhi2020seeing, hood2020prospects}, and recent GCM work implies that clouds may be sequestered at depth in UHJs \citep{Komacek2022}, so the significance of clouds in these data is unclear.

\section{Conclusions} \label{sec:conclusions}

In this work, we have explored the role of jet strength on phase-resolved transmission spectroscopy of UHJs. Although these measurements have an inherent level of degeneracy (due to some physical mechanisms manifesting similarly, see \citealt{Savel2023}), the equatorial jet strength substantially shapes the resulting net Doppler shifts. We offer the following summary points to guide the interpretation of these measurements: 
\begin{itemize}
    \item The rotation of the planet is the most substantial component of the overall Doppler shift and causes a net blueward shift as transit progresses. This is in agreement with previously published observations and atmospheric modeling work \citep{Wardenier2021, 2022Savel,Beltz2023, Pelletier2023,Simonnin2025}. 
    \item  The jet's speed directly alters the slope of the net Doppler shift during ingress and egress.  Faster jets impart a stronger blueward slope than models with weaker or no jets present. As a secondary effect, the jet additionally imparts a blueward slope during mid-transit due to projection effects i.e., due to the measured velocity field being biased toward the hot, inflated portion of the atmosphere. 
    \item This behavior is distinct from that of day--night winds, which generally introduce an offset to phase-resolved Doppler shifts due to their symmetry across the limbs. Thus, the slopes seen in these models' Doppler shifts are primarily due to planetary rotation. 
    \item In addition to having smaller overall net Doppler shifts, the magnetically active models cross the red-to-blueshift threshold significantly later in transit than the drag-free cases. This is due to the asymmetric jet structure and lower overall jet speeds. The weaker jet speed results in a more shallow ingress slope, which results in a later net blueshift.
    \item Jets also impact the width of CCFs. Models with more velocity dispersion (i.e., faster jets) have greater CCF widths, and those with more thermal asymmetry also have slopes in the CCF width over phase.
    \item Semi-analytical work implies that, to leading order, the CCF amplitude does not encode much information about atmospheric asymmetry. However, the CCF shift and width each contain information about the planets' thermal and dynamical inhomogeneities.
\end{itemize}

Multiple promising avenues exist for phase-resolved spectroscopy at high spectral resolution. For instance, HRCCS retrievals using more flexible velocity parameterizations (extending beyond deviations from a circular orbit; i.e., beyond $\Delta K_p$ and $\Delta V_{\rm sys}$, and a constant broadening kernel) may reveal and quantify thermal, dynamical, and chemical variations hidden in extant data. Our work shows that combining multiple transits may yield informative constraints on jet dynamical processes, provided that spectra across multiple nights stack coherently. It is critical that these measurements include ingress and egress, as mid-transit Doppler shifts can be similar between different models. These results may also be applicable to phase-resolved emission spectroscopy, which we will investigate in detail in a later work. 

Of course, the most striking near-future development for phase-resolved spectroscopy will be the Extremely Large Telescopes \citep[ELTs;][]{gilmozzi2007european,johns2012giant}. These facilities, with their substantial increase in collecting area, will enable up to a fourfold increase in velocity precision over today's spectra \citep{dragomir2019characterizing}. Given that the strength of atmospheric jets' impact on our simulations approaches the precision of a single ESPRESSO transit, and that ingress and egress spectroscopy are primarily limited by signal-to-noise considerations, it is clear that the ELTs will constrain the details of atmospheric dynamics \citep{palle2025ground}. This decrease in photon noise will also enable the ELTs to spectrally resolve between 4--5 times as many individual spectral lines as current instruments. Precise measurements of hot Jupiters' vertical velocity structure \citep{Kempton2012,KesseliBeltz2024,Seidel2025} as a function of orbital phase \textit{with a single species} will therefore be possible, providing tight constraints on dynamical processes. Finally, the ELTs' mirrors will be sensitive to fainter stars in a wider volume of space, increasing the number of exoplanets amenable to HRCCS by an order of magnitude \citep{palle2025ground}. This considerable growth in the observable hot Jupiter population will enable comparative exoplanetary science at scale from the ground---and perhaps reveal the drivers and limiters of atmospheric circulation in hot gas giants.    

\begin{acknowledgments}
HB and ABS thank Eliza M.-R. Kempton and Thaddeus D. Komacek for both their insightful comments toward this manuscript and their mentorship over many years. The authors also thank the referee for their thoughtful comments that greatly improved the quality of this work. 

The authors acknowledge the University of Maryland supercomputing resources (\url{http://hpcc.umd.edu}) made available for conducting the research reported in this paper.

This research has made use of the Astrophysics Data System, funded by NASA under Cooperative Agreement 80NSSC21M0056.
\end{acknowledgments}

\software{\texttt{astropy} \citep{astropy:2013, astropy:2018, astropy:2022}, \texttt{FastChem} \citep{Fastchem2018, stock2022fastchem}, \texttt{GitHub Copilot} \citep{chen2021evaluating}, \texttt{JAX} \citep{jax2018github}, \texttt{Jupyter} \citep{granger2021jupyter}, \texttt{Matplotlib} \citep{Hunter:2007}, \texttt{Numpy} \citep{harris2020array}, \texttt{pandas} \citep{mckinney-proc-scipy-2010, reback2020pandas}, \texttt{scipy} \citep{2020SciPy-NMeth},
\texttt{scope} \citep{Savel2025},
\texttt{tqdm} \citep{da2019tqdm}, \texttt{Windspharm} \citep{dawson2016windspharm}}

\restartappendixnumbering

\begin{contribution}

All authors contributed equally to this work. 


\end{contribution}

%

\appendix
\section{Derivations of CCF quantities}\label{appendix:derivation}

\subsection{Amplitude}
Consider a simplifying situation in which an exoplanet's atmosphere is vertically and latitudinally isothermal (at $T(\theta)$), that it is probed across $n$ scale heights, and that gravity $g$ and mean molecular weight $\mu$ are constant throughout the whole observable atmosphere. Then the area of each limb is, expanding as, e.g., \cite{louie2018simulated}:

\begin{equation}
    A({\theta}) = (\pi/2) (R_p + nH(\theta))^2 - (\pi/2)R_p^2 \approx \pi R_p nH(\theta),
\end{equation}
so long as $nH(\theta) << R_p$.

Substituting in the scale height, we have

\begin{equation}
    \begin{split}
        A(\theta) = \frac{n\pi k_BT_{m,e}R_p}{\mu g}. \\
    \end{split}
\end{equation}

Now assume a CCF that is sensitive to line strength or a formally derived likelihood function \citep[e.g.,][]{Brogi2019,Gibson2020}. Then this function, $f$, is sensitive to the amplitude of spectral lines in the transmission spectrum, not only their location. Ignoring questions of model--template mismatch (e.g., sidestepping the question of $f$ degrading due to a 1D model not capturing the line profile of a 3D atmosphere; e.g., \citealt{Flowers2019,Beltz2021}), the strength of the spectral features is proportional to the total occulting area of the exoplanet's atmosphere:

\begin{equation}
    f \propto A.
\end{equation}

Let us parameterize the temperature field as a sinusoid that only varies in longitude:

\begin{equation}
    T = T_0 + \Delta T\cos(\theta - \theta_{offset} - \varphi),
\end{equation}
where $T_0$ is the nightside temperature, $\Delta T$ is the day--night contrast, $\theta$ is the longitude, $\theta_{offset}$ is some offset in the temperature field (e.g., due to Doppler-shifting of the planetary-scale standing wave pattern by the equatorial jet; \citealt{hammond2018wave}), and $\varphi$ is the orbital phase (0 at mid-transit).

With this sinusoidal approximation, we have:

\begin{equation}
    \begin{split}
        f \propto A_m + A_e \\
        f \propto T_0 + \Delta T\cos\bigg{(}\frac{3\pi}{2} -\theta_{offset} - \phi\bigg{)} + T_0 + \Delta T\cos\bigg{(}\frac{\pi}{2} -\theta_{offset} - \phi\bigg{)} \\
        f \propto T_0 + \Delta T\cos\bigg{(}\frac{3\pi}{2} -\theta_{offset} - \phi\bigg{)} + T_0 + \Delta T\cos\bigg{(}\frac{\pi}{2} -\theta_{offset} - \phi\bigg{)} \\
        f \propto 2T_0 + \Delta T(\cos\bigg{(}\frac{3\pi}{2} -\theta_{offset} - \phi\bigg{)} + \cos\bigg{(}\frac{\pi}{2} -\theta_{offset} - \phi\bigg{)}) \\
        f \propto 2T_0 + 2\Delta T(\cos(0.5\bigg{(}\frac{3\pi}{2} -\theta_{offset} - \phi + \frac{\pi}{2} -\theta_{offset} - \phi\bigg{)}))\cos(0.5\bigg{(}\frac{3\pi}{2} -\theta_{offset} - \phi - \frac{\pi}{2} -\theta_{offset}  \phi\bigg{)})) \\
        f \propto 2T_0 + 2\Delta T(\cos(\pi -\theta_{offset} - \phi))\cos\bigg{(}\frac{\pi}{2}\bigg{)}) \\
        f \propto 2T_0 - 2\Delta T(\cos(\theta_{offset} + \phi))\cos\bigg{(}\frac{\pi}{2}\bigg{)}) \\
        f \propto 2T_0.
    \end{split}
\end{equation}
Thus, the total area --- and therefore the strength of the CCF signal --- is conserved over the course of transit under the sinusoidal temperature approximation. That is, the amplitude does \textit{not} encode atmospheric asymmetry here. 

It is worth noting that the mean molecular weight is not necessarily constant over the full atmosphere. Strong gradients in mean molecular weight can emerge in atmospheres in regions of substantial molecular dissociation \citep[e.g.,][]{parmentier2018,Tan_2019}. We expect that departures from this leading-order behavior would therefore be particularly salient for the ultrahot Jupiter population \citep{Lothringer2018,Bell_2018,2019Arcangeli}.

\subsection{Net Doppler shift}

A reasonable interpretation is that the changing Doppler shift over transit is predominantly due to a smooth change in the visible thermal structure, as a continuously changing temperature field rotates through an observability window \citep[e.g,.][]{Wardenier2021}. In this case we have:

\begin{equation}
    \begin{split}
        v_{net} = \frac{A_m(v_m + v_{rot}) + A_e(-v_e - v_{rot})}{A_m + A_e}, \\
    \end{split}
\end{equation}
where the evening limb velocities in the planet frame are assigned a negative value, as they are projected \textit{toward} the observer.

 Maintaining the previously described arguments regarding constant mean molecular weight and gravity, and retaining the result for area from the amplitude derivation, we then have:

\begin{equation}
    \begin{split}
        v_{net} = \frac{T_m(v_m + v_{rot}) + T_e(-v_e - v_{rot})}{2T_0}. \\
\end{split}
\end{equation}
Then        
\begin{equation}
\begin{split}    
        v_{net} = \frac{T_mv_m - T_ev_e + v_{rot}(T_m - T_e)}{2T_0} \\
        v_{net} = (\frac{1}{2T_0})((T_0 + \Delta T\cos(3\pi/2 - \theta_{offset} - \varphi))v_m - (T_0 + \Delta T\cos(\pi/2 - \theta_{offset} - \varphi))v_e \\+ \Delta Tv_{rot}(\cos(3\pi/2 - \theta_{offset} - \varphi) - \cos(\pi/2 - \theta_{offset} - \varphi))) \\
        v_{net} = (\frac{1}{2T_0})((T_0 + \Delta T\cos(3\pi/2 - \theta_{offset} - \varphi))v_m - (T_0 + \Delta T\cos(\pi/2 - \theta_{offset} - \varphi))_ev_e \\+ \Delta Tv_{rot}(-2\sin(0.5(3\pi/2 - \theta_{offset} - \varphi + \pi/2 - \theta_{offset} - \varphi))\sin(0.5((3\pi/2 - \theta_{offset} - \varphi - \pi/2 + \theta_{offset} + \varphi)) \\
        v_{net} = (\frac{1}{2T_0})((T_0 + \Delta T\cos(3\pi/2 - \theta_{offset} - \varphi))v_m - (T_0 + \Delta T\cos(\pi/2 - \theta_{offset} - \varphi))v_e \\+ \Delta Tv_{rot}(-2\sin(\pi - \theta_{offset} - \varphi)\sin(\pi/2) \\
        v_{net} = (\frac{1}{2T_0})((T_0 + \Delta T\cos(3\pi/2 - \theta_{offset} - \varphi))v_m - (T_0 + \Delta T\cos(\pi/2 - \theta_{offset} - \varphi))v_e \\- 2\Delta Tv_{rot}\sin( \theta_{offset} - \varphi) \\
        v_{net} = (\frac{1}{2T_0})((T_0 - \Delta T\sin(\theta_{offset} + \varphi))v_m - (T_0 + \Delta T\sin(\theta_{offset} + \varphi))v_e - 2\Delta Tv_{rot}\sin( \theta_{offset} + \varphi) \\
                v_{net} = (\frac{1}{2T_0})(T_0v_m - \Delta T\sin(\theta_{offset} + \varphi)v_m - T_0v_e - \Delta T\sin(\theta_{offset} + \varphi)v_e - 2\Delta Tv_{rot}\sin( \theta_{offset} + \varphi) \\
        v_{net} = (\frac{1}{2T_0})(T_0v_m - T_0v_e - \Delta T\sin(\theta_{offset} + \varphi)(-v_m -v_e - 2v_{rot})) \\
        v_{net} = (\frac{v_m - v_e}{2}) -\frac{\Delta T}{2T_0}\sin(\theta_{offset} + \varphi)(-v_m -v_e - 2v_{rot}). \\
    \end{split}
\end{equation}

This is the net velocity. Now taking the derivative with respect to phase, only the sinusoid term survives:

\begin{equation}
    \begin{split}
        \frac{d(v_{net})}{d\phi} = \bigg{(}\frac{2v_{rot} + v_m + v_e}{2T_0}\bigg{)}\cos(\theta_{offset} + \phi)\Delta T.
    \end{split}
\end{equation}

Altogether, the slope of Doppler shifts during transit is set by both thermal and velocity gradients. Notably, a gradient in either field leads to zero slope. In practice, rotation always provides a gradient in the projected velocity field (slight as it may be), so only an additional thermal gradient is required to produce a slope. 

\subsection{Width}
First, let us consider different contributors to the width of the CCF. Approximating these sources as Gaussian distributions, the full width of the CCF goes as:

\begin{figure}
    \centering
    \includegraphics[width=0.8\linewidth]{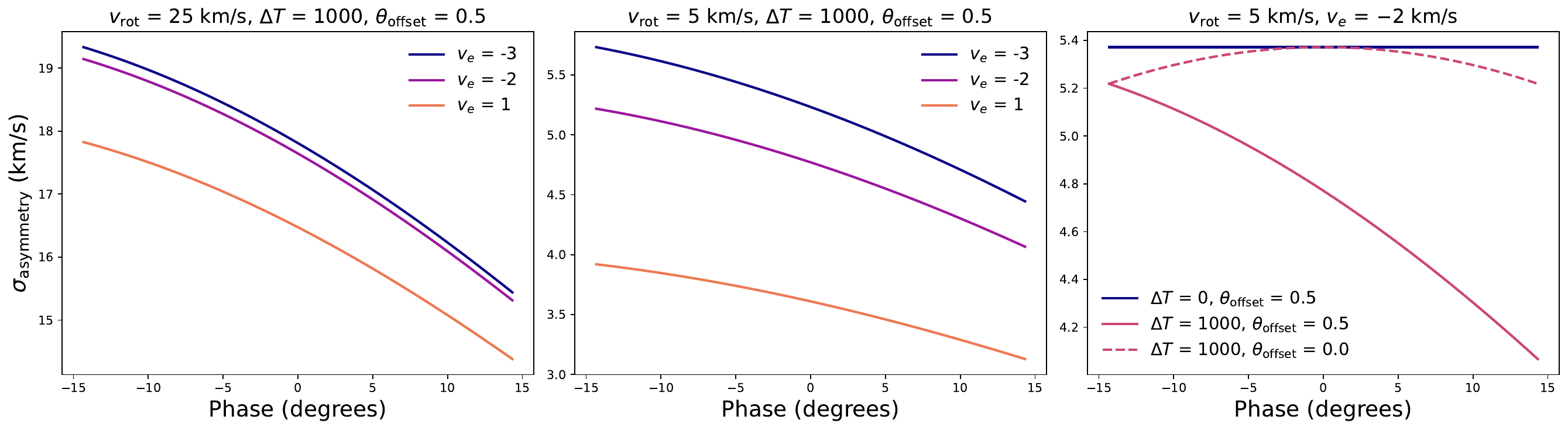}
    \caption{Evaluating Eq.~\ref{eq:ccf_width}--Eq.~\ref{eq:variance} as a function of phase. Left: changing the velocity asymmetry (by altering the projected evening limb velocity). Middle: same as the left, but with a decreased rotational velocity. Right: changing only the thermal structure (its day--night contrast $\Delta T$ and its offset $\theta_{offset}$). Note that this broadening does \textit{not} include contributions from all sources of line broadening---only the broadening due to macroscopic atmospheric asymmetry.}
    \label{fig:asymmetry_eq}
\end{figure}

\begin{equation}
    \sigma^2_{tot} \approx \sigma^2_{natural} + \sigma^2_{alias} + \sigma^2_{thermal}+ \sigma^2_{pressure} + \sigma^2_{asymmetry},
\end{equation}
where $\sigma^2_{natural}$ is the ``natural'' broadening due to quantum uncertainty, $\sigma^2_{alias}$ is due to aliasing of a single-species cross-correlation template against that of spectral species, $\sigma^2_{thermal}$ is \textit{microscopic} thermal broadening of the spectral lines of interest, $\sigma^2_{pressure}$ is due to pressure broadening, and $\sigma^2_{asymmetry}$ is due to \textit{macroscopic} differences in thermal and velocity structure. Given that these terms are additive if we assume them to be Gaussian-like, the leading-order evolution of the CCF broadening term as a function of phase is simply

\begin{equation}
    \frac{d(\sigma_{tot}^2)}{d\varphi} \approx \frac{d\sigma^2_{asymmetry}}{d\varphi}
\end{equation}

What is the broadening parameter due to macroscopic asymmetry? The distribution of velocities due to solid-body rotation is well known for transmission spectroscopy \citep[e.g.,][]{Gandhi2022}. For morning and evening limbs with constant velocities, these solid-body rotation distributions are simply shifted by that constant velocity. The distribution of velocities is then:

\begin{equation}\label{eq:ccf_width}
    \begin{split}
        P(v_z) \propto\frac{A_e}{\sqrt{v^2_{rot} - (v_z - v_e)^2}}\mathbb{1}_{-v_{rot} + v_e < v_z < v_e} + \frac{A_m}{\sqrt{v^2_{rot} - (v_z - v_m)^2}}\mathbb{1}_{v_m < v_z < v_{rot} + v_m} \\
        P(v_z) \propto\frac{T_0 + \Delta T\cos(\frac{\pi}{2} - \theta_{offset} - \phi)}{\sqrt{v^2_{rot} - (v_z - v_e)^2}}\mathbb{1}_{-v_{rot} + v_e < v_z < v_e} + \frac{T_0 + \Delta T\cos(\frac{3\pi}{2} - \theta_{offset} - \phi)}{\sqrt{v^2_{rot} - (v_z - v_m)^2}}\mathbb{1}_{v_m < v_z < v_{rot} + v_m},
    \end{split}
\end{equation}

where the indicator function $\mathbb{1}$ controls the bounds over which each term is defined. To calculate the broadening from this probability function, we note:

\begin{equation}\label{eq:variance}
\begin{split}
    <v_z> = \int v_z P(v_z)dz \\
    <v_z^2> = \int v_z^2 P(v_z)dz \\
    \sigma_{asymmetry}^2 = <v_z^2> -  <v_z>.
\end{split}
\end{equation}

These expressions are challenging to evaluate with traditional methods and integral calculators. We therefore numerically integrate to find $\sigma_{asymmetry}^2$ as a function of orbital phase, changing one quantity at a time, in Fig.~\ref{fig:asymmetry_eq}. In the absence of any thermal inhomogeneities and in a symmetric wind field, there is no change in the broadening parameter. If the thermal structure is inhomogeneous but symmetric about the substellar point, the CCF width evolves like a parabola. If the thermal structure exhibits a hotspot offset, then there is a somewhat linear slope in $\sigma_{asymmetry}^2$ through transit. Given a set thermal inhomogeneity, the strength of the slope can increase with the difference in wind speed between the limbs. This behavior is made especially obvious when the rotational velocity is drastically increased.

\section{Sensitivity of results to barotropic extension}\label{appendix:barotropic_test}
To assess how much our results change under reasonable perturbations to our barotropically extended wind profile, we conduct a sensitivity test to the wind field at pressures less than 10~$\mu$bar. To do so, we test the difference between phase-resolved Doppler shifts calculated with the barotropic extension and those calculated with a constant multiplier of 1.5 to the barotropic extension.

\begin{figure}
    \centering
    \includegraphics[width=0.5\linewidth]{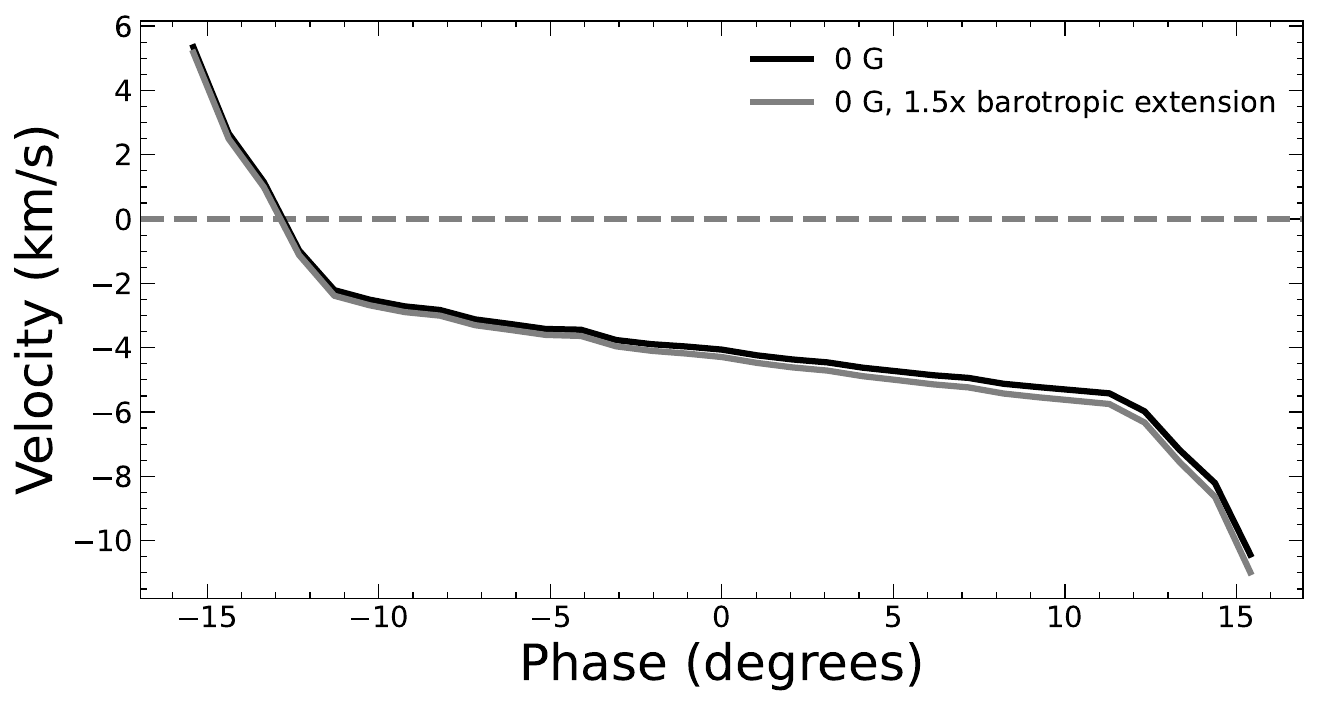}
    \caption{The results of our sensitivity test to the barotropic extension. Altering the extended velocity profile by a factor of 1.5 only minorly alters our calculated phase-resolved Doppler shifts.}
    \label{fig:barotropic_test}
\end{figure}

The results (Fig~\ref{fig:barotropic_test}) imply that our phase-resolved Doppler shifts change by less than 1~km\,s$^{-1}$. upon making this perturbation. We therefore conclude that our work is not dependent on the exact velocity profile at the lowest pressures, likely due to the small number of lines that probe these pressures.

\bibliography{sample7}{}

@article{malik2019self,
  title={Self-luminous and irradiated exoplanetary atmospheres explored with HELIOS},
  author={Malik, Matej and Kitzmann, Daniel and Mendon{\c{c}}a, Jo{\~a}o M and Grimm, Simon L and Marleau, Gabriel-Dominique and Linder, Esther F and Tsai, Shang-Min and Heng, Kevin},
  journal={The Astronomical Journal},
  volume={157},
  number={5},
  pages={170},
  year={2019},
  publisher={IOP Publishing}
}

@article{nortmann2024crires,
  title={CRIRES transmission spectroscopy of WASP-127 b},
  author={Nortmann, L and Lesjak, F and Yan, F and Cont, D and Czesla, S and Lavail, A and Rains, AD and Nagel, E and Boldt-Christmas, L and Hatzes, A and others},
  journal={Detection of the resolved signatures of a supersonic equatorial jet and cool poles in a hot planet, publication Title: arXiv. org},
  year={2024}
}

@article{molliere2019petitradtrans,
  title={petitRADTRANS-A Python radiative transfer package for exoplanet characterization and retrieval},
  author={Molli{\`e}re, P and Wardenier, JP and Van Boekel, R and Henning, Th and Molaverdikhani, K and Snellen, IAG},
  journal={Astronomy \& Astrophysics},
  volume={627},
  pages={A67},
  year={2019},
  publisher={EDP Sciences}
}

@article{brown2001transmission,
  title={Transmission spectra as diagnostics of extrasolar giant planet atmospheres},
  author={Brown, Timothy M},
  journal={The Astrophysical Journal},
  volume={553},
  number={2},
  pages={1006--1026},
  year={2001}
}

@article{louie2018simulated,
  title={Simulated JWST/NIRISS transit spectroscopy of anticipated tess planets compared to select discoveries from space-based and ground-based surveys},
  author={Louie, Dana R and Deming, Drake and Albert, Loic and Bouma, LG and Bean, Jacob and Lopez-Morales, Mercedes},
  journal={Publications of the Astronomical Society of the Pacific},
  volume={130},
  number={986},
  pages={044401},
  year={2018},
  publisher={The Astronomical Society of the Pacific}
}

@article{hammond2018wave,
  title={Wave-mean flow interactions in the atmospheric circulation of tidally locked planets},
  author={Hammond, Mark and Pierrehumbert, Raymond T},
  journal={The Astrophysical Journal},
  volume={869},
  number={1},
  pages={65},
  year={2018},
  publisher={The American Astronomical Society}
}

@article{parmentier2021cloudy,
  title={The cloudy shape of hot Jupiter thermal phase curves},
  author={Parmentier, Vivien and Showman, Adam P and Fortney, Jonathan J},
  journal={Monthly Notices of the Royal Astronomical Society},
  volume={501},
  number={1},
  pages={78--108},
  year={2021},
  publisher={Oxford University Press}
}

@article{hood2020prospects,
  title={Prospects for characterizing the haziest sub-Neptune exoplanets with high-resolution spectroscopy},
  author={Hood, Callie E and Fortney, Jonathan J and Line, Michael R and Martin, Emily C and Morley, Caroline V and Birkby, Jayne L and Rustamkulov, Zafar and Lupu, Roxana E and Freedman, Richard S},
  journal={The Astronomical Journal},
  volume={160},
  number={5},
  pages={198},
  year={2020},
  publisher={IOP Publishing}
}

@article{gandhi2020seeing,
  title={Seeing above the clouds with high-resolution spectroscopy},
  author={Gandhi, Siddharth and Brogi, Matteo and Webb, Rebecca K},
  journal={Monthly Notices of the Royal Astronomical Society},
  volume={498},
  number={1},
  pages={194--204},
  year={2020},
  publisher={Oxford University Press}
}

@ARTICLE{2020SciPy-NMeth,
  author  = {Virtanen, Pauli and Gommers, Ralf and Oliphant, Travis E. and
            Haberland, Matt and Reddy, Tyler and Cournapeau, David and
            Burovski, Evgeni and Peterson, Pearu and Weckesser, Warren and
            Bright, Jonathan and {van der Walt}, St{\'e}fan J. and
            Brett, Matthew and Wilson, Joshua and Millman, K. Jarrod and
            Mayorov, Nikolay and Nelson, Andrew R. J. and Jones, Eric and
            Kern, Robert and Larson, Eric and Carey, C J and
            Polat, {\.I}lhan and Feng, Yu and Moore, Eric W. and
            {VanderPlas}, Jake and Laxalde, Denis and Perktold, Josef and
            Cimrman, Robert and Henriksen, Ian and Quintero, E. A. and
            Harris, Charles R. and Archibald, Anne M. and
            Ribeiro, Ant{\^o}nio H. and Pedregosa, Fabian and
            {van Mulbregt}, Paul and {SciPy 1.0 Contributors}},
  title   = {{{SciPy} 1.0: Fundamental Algorithms for Scientific
            Computing in Python}},
  journal = {Nature Methods},
  year    = {2020},
  volume  = {17},
  pages   = {261--272},
  adsurl  = {https://rdcu.be/b08Wh},
  doi     = {10.1038/s41592-019-0686-2},
}

@article{da2019tqdm,
  title={tqdm: A Fast, Extensible Progress Meter for Python and CLI},
  author={da Costa-Luis, Casper},
  journal={Journal of Open Source Software},
  volume={4},
  number={37},
  pages={1277},
  year={2019}
}

@software{reback2020pandas,
    author       = {The pandas development team},
    title        = {pandas-dev/pandas: Pandas},
    month        = feb,
    year         = 2020,
    publisher    = {Zenodo},
    version      = {latest},
    doi          = {10.5281/zenodo.3509134},
    url          = {https://doi.org/10.5281/zenodo.3509134}
}

@InProceedings{ mckinney-proc-scipy-2010,
  author    = { {W}es {M}c{K}inney },
  title     = { {D}ata {S}tructures for {S}tatistical {C}omputing in {P}ython },
  booktitle = { {P}roceedings of the 9th {P}ython in {S}cience {C}onference },
  pages     = { 56 - 61 },
  year      = { 2010 },
  editor    = { {S}t\'efan van der {W}alt and {J}arrod {M}illman },
  doi       = { 10.25080/Majora-92bf1922-00a }
}

@software{jax2018github,
  author = {James Bradbury and Roy Frostig and Peter Hawkins and Matthew James Johnson and Chris Leary and Dougal Maclaurin and George Necula and Adam Paszke and Jake Vander{P}las and Skye Wanderman-{M}ilne and Qiao Zhang},
  title = {{JAX}: composable transformations of {P}ython+{N}um{P}y programs},
  url = {http://github.com/jax-ml/jax},
  version = {0.3.13},
  year = {2018},
}

@Article{         harris2020array,
 title         = {Array programming with {NumPy}},
 author        = {Charles R. Harris and K. Jarrod Millman and St{\'{e}}fan J.
                 van der Walt and Ralf Gommers and Pauli Virtanen and David
                 Cournapeau and Eric Wieser and Julian Taylor and Sebastian
                 Berg and Nathaniel J. Smith and Robert Kern and Matti Picus
                 and Stephan Hoyer and Marten H. van Kerkwijk and Matthew
                 Brett and Allan Haldane and Jaime Fern{\'{a}}ndez del
                 R{\'{i}}o and Mark Wiebe and Pearu Peterson and Pierre
                 G{\'{e}}rard-Marchant and Kevin Sheppard and Tyler Reddy and
                 Warren Weckesser and Hameer Abbasi and Christoph Gohlke and
                 Travis E. Oliphant},
 year          = {2020},
 month         = sep,
 journal       = {Nature},
 volume        = {585},
 number        = {7825},
 pages         = {357--362},
 doi           = {10.1038/s41586-020-2649-2},
 publisher     = {Springer Science and Business Media {LLC}},
 url           = {https://doi.org/10.1038/s41586-020-2649-2}
}

@Article{Hunter:2007,
  Author    = {Hunter, J. D.},
  Title     = {Matplotlib: A 2D graphics environment},
  Journal   = {Computing in Science \& Engineering},
  Volume    = {9},
  Number    = {3},
  Pages     = {90--95},
  publisher = {IEEE COMPUTER SOC},
  doi       = {10.1109/MCSE.2007.55},
  year      = 2007
}

@article{granger2021jupyter,
  title={Jupyter: Thinking and storytelling with code and data},
  author={Granger, Brian E and P{\'e}rez, Fernando},
  journal={Computing in Science \& Engineering},
  volume={23},
  number={2},
  pages={7--14},
  year={2021},
  publisher={IEEE}
}

@article{stock2022fastchem,
  title={FastChem 2: an improved computer program to determine the gas-phase chemical equilibrium composition for arbitrary element distributions},
  author={Stock, Joachim W and Kitzmann, Daniel and Patzer, A Beate C},
  journal={Monthly Notices of the Royal Astronomical Society},
  volume={517},
  number={3},
  pages={4070--4080},
  year={2022},
  publisher={Oxford University Press}
}

@article{astropy:2013,
Adsurl = {http://adsabs.harvard.edu/abs/2013A%26A...558A..33A},
Archiveprefix = {arXiv},
Author = {{Astropy Collaboration} and {Robitaille}, T.~P. and {Tollerud}, E.~J. and {Greenfield}, P. and {Droettboom}, M. and {Bray}, E. and {Aldcroft}, T. and {Davis}, M. and {Ginsburg}, A. and {Price-Whelan}, A.~M. and {Kerzendorf}, W.~E. and {Conley}, A. and {Crighton}, N. and {Barbary}, K. and {Muna}, D. and {Ferguson}, H. and {Grollier}, F. and {Parikh}, M.~M. and {Nair}, P.~H. and {Unther}, H.~M. and {Deil}, C. and {Woillez}, J. and {Conseil}, S. and {Kramer}, R. and {Turner}, J.~E.~H. and {Singer}, L. and {Fox}, R. and {Weaver}, B.~A. and {Zabalza}, V. and {Edwards}, Z.~I. and {Azalee Bostroem}, K. and {Burke}, D.~J. and {Casey}, A.~R. and {Crawford}, S.~M. and {Dencheva}, N. and {Ely}, J. and {Jenness}, T. and {Labrie}, K. and {Lim}, P.~L. and {Pierfederici}, F. and {Pontzen}, A. and {Ptak}, A. and {Refsdal}, B. and {Servillat}, M. and {Streicher}, O.},
Doi = {10.1051/0004-6361/201322068},
Eid = {A33},
Eprint = {1307.6212},
Journal = {\aap},
Month = oct,
Pages = {A33},
Primaryclass = {astro-ph.IM},
Title = {{Astropy: A community Python package for astronomy}},
Volume = 558,
Year = 2013}

@ARTICLE{astropy:2018,
       author = {{Astropy Collaboration} and {Price-Whelan}, A.~M. and
         {Sip{\H{o}}cz}, B.~M. and {G{\"u}nther}, H.~M. and {Lim}, P.~L. and
         {Crawford}, S.~M. and {Conseil}, S. and {Shupe}, D.~L. and
         {Craig}, M.~W. and {Dencheva}, N. and {Ginsburg}, A. and {Vand
        erPlas}, J.~T. and {Bradley}, L.~D. and {P{\'e}rez-Su{\'a}rez}, D. and
         {de Val-Borro}, M. and {Aldcroft}, T.~L. and {Cruz}, K.~L. and
         {Robitaille}, T.~P. and {Tollerud}, E.~J. and {Ardelean}, C. and
         {Babej}, T. and {Bach}, Y.~P. and {Bachetti}, M. and {Bakanov}, A.~V. and
         {Bamford}, S.~P. and {Barentsen}, G. and {Barmby}, P. and
         {Baumbach}, A. and {Berry}, K.~L. and {Biscani}, F. and {Boquien}, M. and
         {Bostroem}, K.~A. and {Bouma}, L.~G. and {Brammer}, G.~B. and
         {Bray}, E.~M. and {Breytenbach}, H. and {Buddelmeijer}, H. and
         {Burke}, D.~J. and {Calderone}, G. and {Cano Rodr{\'\i}guez}, J.~L. and
         {Cara}, M. and {Cardoso}, J.~V.~M. and {Cheedella}, S. and {Copin}, Y. and
         {Corrales}, L. and {Crichton}, D. and {D'Avella}, D. and {Deil}, C. and
         {Depagne}, {\'E}. and {Dietrich}, J.~P. and {Donath}, A. and
         {Droettboom}, M. and {Earl}, N. and {Erben}, T. and {Fabbro}, S. and
         {Ferreira}, L.~A. and {Finethy}, T. and {Fox}, R.~T. and
         {Garrison}, L.~H. and {Gibbons}, S.~L.~J. and {Goldstein}, D.~A. and
         {Gommers}, R. and {Greco}, J.~P. and {Greenfield}, P. and
         {Groener}, A.~M. and {Grollier}, F. and {Hagen}, A. and {Hirst}, P. and
         {Homeier}, D. and {Horton}, A.~J. and {Hosseinzadeh}, G. and {Hu}, L. and
         {Hunkeler}, J.~S. and {Ivezi{\'c}}, {\v{Z}}. and {Jain}, A. and
         {Jenness}, T. and {Kanarek}, G. and {Kendrew}, S. and {Kern}, N.~S. and
         {Kerzendorf}, W.~E. and {Khvalko}, A. and {King}, J. and {Kirkby}, D. and
         {Kulkarni}, A.~M. and {Kumar}, A. and {Lee}, A. and {Lenz}, D. and
         {Littlefair}, S.~P. and {Ma}, Z. and {Macleod}, D.~M. and
         {Mastropietro}, M. and {McCully}, C. and {Montagnac}, S. and
         {Morris}, B.~M. and {Mueller}, M. and {Mumford}, S.~J. and {Muna}, D. and
         {Murphy}, N.~A. and {Nelson}, S. and {Nguyen}, G.~H. and
         {Ninan}, J.~P. and {N{\"o}the}, M. and {Ogaz}, S. and {Oh}, S. and
         {Parejko}, J.~K. and {Parley}, N. and {Pascual}, S. and {Patil}, R. and
         {Patil}, A.~A. and {Plunkett}, A.~L. and {Prochaska}, J.~X. and
         {Rastogi}, T. and {Reddy Janga}, V. and {Sabater}, J. and
         {Sakurikar}, P. and {Seifert}, M. and {Sherbert}, L.~E. and
         {Sherwood-Taylor}, H. and {Shih}, A.~Y. and {Sick}, J. and
         {Silbiger}, M.~T. and {Singanamalla}, S. and {Singer}, L.~P. and
         {Sladen}, P.~H. and {Sooley}, K.~A. and {Sornarajah}, S. and
         {Streicher}, O. and {Teuben}, P. and {Thomas}, S.~W. and
         {Tremblay}, G.~R. and {Turner}, J.~E.~H. and {Terr{\'o}n}, V. and
         {van Kerkwijk}, M.~H. and {de la Vega}, A. and {Watkins}, L.~L. and
         {Weaver}, B.~A. and {Whitmore}, J.~B. and {Woillez}, J. and
         {Zabalza}, V. and {Astropy Contributors}},
        title = "{The Astropy Project: Building an Open-science Project and Status of the v2.0 Core Package}",
      journal = {\aj},
         year = 2018,
        month = sep,
       volume = {156},
       number = {3},
          eid = {123},
        pages = {123},
          doi = {10.3847/1538-3881/aabc4f},
archivePrefix = {arXiv},
       eprint = {1801.02634},
 primaryClass = {astro-ph.IM},
       adsurl = {https://ui.adsabs.harvard.edu/abs/2018AJ....156..123A}
}

@ARTICLE{astropy:2022,
       author = {{Astropy Collaboration} and {Price-Whelan}, Adrian M. and {Lim}, Pey Lian and {Earl}, Nicholas and {Starkman}, Nathaniel and {Bradley}, Larry and {Shupe}, David L. and {Patil}, Aarya A. and {Corrales}, Lia and {Brasseur}, C.~E. and {N{"o}the}, Maximilian and {Donath}, Axel and {Tollerud}, Erik and {Morris}, Brett M. and {Ginsburg}, Adam and {Vaher}, Eero and {Weaver}, Benjamin A. and {Tocknell}, James and {Jamieson}, William and {van Kerkwijk}, Marten H. and {Robitaille}, Thomas P. and {Merry}, Bruce and {Bachetti}, Matteo and {G{"u}nther}, H. Moritz and {Aldcroft}, Thomas L. and {Alvarado-Montes}, Jaime A. and {Archibald}, Anne M. and {B{'o}di}, Attila and {Bapat}, Shreyas and {Barentsen}, Geert and {Baz{'a}n}, Juanjo and {Biswas}, Manish and {Boquien}, M{'e}d{'e}ric and {Burke}, D.~J. and {Cara}, Daria and {Cara}, Mihai and {Conroy}, Kyle E. and {Conseil}, Simon and {Craig}, Matthew W. and {Cross}, Robert M. and {Cruz}, Kelle L. and {D'Eugenio}, Francesco and {Dencheva}, Nadia and {Devillepoix}, Hadrien A.~R. and {Dietrich}, J{"o}rg P. and {Eigenbrot}, Arthur Davis and {Erben}, Thomas and {Ferreira}, Leonardo and {Foreman-Mackey}, Daniel and {Fox}, Ryan and {Freij}, Nabil and {Garg}, Suyog and {Geda}, Robel and {Glattly}, Lauren and {Gondhalekar}, Yash and {Gordon}, Karl D. and {Grant}, David and {Greenfield}, Perry and {Groener}, Austen M. and {Guest}, Steve and {Gurovich}, Sebastian and {Handberg}, Rasmus and {Hart}, Akeem and {Hatfield-Dodds}, Zac and {Homeier}, Derek and {Hosseinzadeh}, Griffin and {Jenness}, Tim and {Jones}, Craig K. and {Joseph}, Prajwel and {Kalmbach}, J. Bryce and {Karamehmetoglu}, Emir and {Ka{l}uszy{'n}ski}, Miko{l}aj and {Kelley}, Michael S.~P. and {Kern}, Nicholas and {Kerzendorf}, Wolfgang E. and {Koch}, Eric W. and {Kulumani}, Shankar and {Lee}, Antony and {Ly}, Chun and {Ma}, Zhiyuan and {MacBride}, Conor and {Maljaars}, Jakob M. and {Muna}, Demitri and {Murphy}, N.~A. and {Norman}, Henrik and {O'Steen}, Richard and {Oman}, Kyle A. and {Pacifici}, Camilla and {Pascual}, Sergio and {Pascual-Granado}, J. and {Patil}, Rohit R. and {Perren}, Gabriel I. and {Pickering}, Timothy E. and {Rastogi}, Tanuj and {Roulston}, Benjamin R. and {Ryan}, Daniel F. and {Rykoff}, Eli S. and {Sabater}, Jose and {Sakurikar}, Parikshit and {Salgado}, Jes{'u}s and {Sanghi}, Aniket and {Saunders}, Nicholas and {Savchenko}, Volodymyr and {Schwardt}, Ludwig and {Seifert-Eckert}, Michael and {Shih}, Albert Y. and {Jain}, Anany Shrey and {Shukla}, Gyanendra and {Sick}, Jonathan and {Simpson}, Chris and {Singanamalla}, Sudheesh and {Singer}, Leo P. and {Singhal}, Jaladh and {Sinha}, Manodeep and {Sip{H{o}}cz}, Brigitta M. and {Spitler}, Lee R. and {Stansby}, David and {Streicher}, Ole and {{{S}}umak}, Jani and {Swinbank}, John D. and {Taranu}, Dan S. and {Tewary}, Nikita and {Tremblay}, Grant R. and {Val-Borro}, Miguel de and {Van Kooten}, Samuel J. and {Vasovi{'c}}, Zlatan and {Verma}, Shresth and {de Miranda Cardoso}, Jos{'e} Vin{'i}cius and {Williams}, Peter K.~G. and {Wilson}, Tom J. and {Winkel}, Benjamin and {Wood-Vasey}, W.~M. and {Xue}, Rui and {Yoachim}, Peter and {Zhang}, Chen and {Zonca}, Andrea and {Astropy Project Contributors}},
        title = "{The Astropy Project: Sustaining and Growing a Community-oriented Open-source Project and the Latest Major Release (v5.0) of the Core Package}",
      journal = {\apj},
         year = 2022,
        month = aug,
       volume = {935},
       number = {2},
          eid = {167},
        pages = {167},
          doi = {10.3847/1538-4357/ac7c74},
archivePrefix = {arXiv},
       eprint = {2206.14220},
 primaryClass = {astro-ph.IM},
       adsurl = {https://ui.adsabs.harvard.edu/abs/2022ApJ...935..167A}
}

@article{johns2012giant,
  title={Giant magellan telescope: overview},
  author={Johns, Matt and McCarthy, Patrick and Raybould, Keith and Bouchez, Antonin and Farahani, Arash and Filgueira, Jose and Jacoby, George and Shectman, Steve and Sheehan, Michael},
  journal={Ground-based and Airborne Telescopes IV},
  volume={8444},
  pages={526--541},
  year={2012},
  publisher={SPIE}
}

@article{gilmozzi2007european,
  title={The European extremely large telescope (E-ELT)},
  author={Gilmozzi, Roberto and Spyromilio, Jason},
  journal={The Messenger},
  volume={127},
  number={11},
  pages={3},
  year={2007}
}

@article{palle2025ground,
  title={Ground-breaking exoplanet science with the ANDES spectrograph at the ELT},
  author={Palle, Enric and Biazzo, Katia and Bolmont, Emeline and Molli{\`e}re, Paul and Poppenhaeger, Katja and Birkby, Jayne and Brogi, Matteo and Chauvin, Gael and Chiavassa, Andrea and Hoeijmakers, Jens and others},
  journal={Experimental Astronomy},
  volume={59},
  number={3},
  pages={1--84},
  year={2025},
  publisher={Springer}
}

@article{dragomir2019characterizing,
  title={Characterizing the Atmospheres of Irradiated Exoplanets at High Spectral Resolution},
  author={Dragomir, Diana and Kempton, Eliza and Bean, Jacob and Crossfield, Ian and Gaidos, Eric and Lewis, Nikole and Line, Michael and Lupu, Roxana and Zhou, George},
  journal={arXiv preprint arXiv:1903.09173},
  year={2019}
}

@article{kempton2014high,
  title={High resolution transmission spectroscopy as a diagnostic for Jovian exoplanet atmospheres: constraints from theoretical models},
  author={Kempton, Eliza M-R and Perna, Rosalba and Heng, Kevin},
  journal={The Astrophysical Journal},
  volume={795},
  number={1},
  pages={24},
  year={2014},
  publisher={IOP Publishing}
}

@article{Showman:2020rev,
	author = {A.P. Showman and X. Tan and V. Parmentier},
	doi = {10.1007/s11214-020-00758-8},
	journal = {Space Science Reviews},
	pages = {139},
	title = {Atmospheric dynamics of hot giant planets and brown dwarfs},
	volume = {216},
	year = {2020}}

@ARTICLE{Beltz2022a,
       author = {{Beltz}, Hayley and {Rauscher}, Emily and {Roman}, Michael T. and {Guilliat}, Abigail},
        title = "{Exploring the Effects of Active Magnetic Drag in a General Circulation Model of the Ultrahot Jupiter WASP-76b}",
      journal = {\aj},
         year = 2022,
        month = jan,
       volume = {163},
       number = {1},
          eid = {35},
        pages = {35},
          doi = {10.3847/1538-3881/ac3746},
archivePrefix = {arXiv},
       eprint = {2109.13371},
 primaryClass = {astro-ph.EP},
       adsurl = {https://ui.adsabs.harvard.edu/abs/2022AJ....163...35B}
      }

@ARTICLE{Beltz2022b,
      author = {{Beltz}, Hayley and {Rauscher}, Emily and {Kempton}, Eliza M. -R. and {Malsky}, Isaac and {Ochs}, Grace and {Arora}, Mireya and {Savel}, Arjun},
        title = "{Magnetic Drag and 3D Effects in Theoretical High-resolution Emission Spectra of Ultrahot Jupiters: the Case of WASP-76b}",
      journal = {\aj},
         year = 2022,
        month = oct,
       volume = {164},
       number = {4},
          eid = {140},
        pages = {140},
          doi = {10.3847/1538-3881/ac897b},
archivePrefix = {arXiv},
       eprint = {2204.12996},
 primaryClass = {astro-ph.EP},
       adsurl = {https://ui.adsabs.harvard.edu/abs/2022AJ....164..140B}
}

@ARTICLE{Beltz2023,
       author = {{Beltz}, Hayley and {Rauscher}, Emily and {Kempton}, Eliza and {Malsky}, Isaac and {Savel}, Arjun},
        title = "{Magnetic Effects and 3D Structure in Theoretical High-Resolution Transmission Spectra of Ultrahot Jupiters: the Case of WASP-76b}",
      journal = {arXiv e-prints},
         year = 2023,
        month = feb,
          eid = {arXiv:2302.13969},
        pages = {arXiv:2302.13969},
          doi = {10.48550/arXiv.2302.13969},
archivePrefix = {arXiv},
       eprint = {2302.13969},
 primaryClass = {astro-ph.EP},
       adsurl = {https://ui.adsabs.harvard.edu/abs/2023arXiv230213969B}
}

@ARTICLE{Gandhi2022,
       author = {{Gandhi}, Siddharth and {Kesseli}, Aurora and {Snellen}, Ignas and {Brogi}, Matteo and {Wardenier}, Joost P. and {Parmentier}, Vivien and {Welbanks}, Luis and {Savel}, Arjun B.},
        title = "{Spatially resolving the terminator: variation of Fe, temperature, and winds in WASP-76 b across planetary limbs and orbital phase}",
      journal = {\mnras},
         year = 2022,
        month = sep,
       volume = {515},
       number = {1},
        pages = {749-766},
          doi = {10.1093/mnras/stac1744},
archivePrefix = {arXiv},
       eprint = {2206.11268},
 primaryClass = {astro-ph.EP},
       adsurl = {https://ui.adsabs.harvard.edu/abs/2022MNRAS.515..749G}
}

@ARTICLE{Kempton2012,
       author = {{Miller-Ricci Kempton}, Eliza and {Rauscher}, Emily},
        title = "{Constraining High-speed Winds in Exoplanet Atmospheres through Observations of Anomalous Doppler Shifts during Transit}",
      journal = {\apj},
         year = 2012,
        month = jun,
       volume = {751},
       number = {2},
          eid = {117},
        pages = {117},
          doi = {10.1088/0004-637X/751/2/117},
archivePrefix = {arXiv},
       eprint = {1109.2270},
 primaryClass = {astro-ph.EP},
       adsurl = {https://ui.adsabs.harvard.edu/abs/2012ApJ...751..117M}
}

@ARTICLE{2012RauscherGCM,
       author = {{Rauscher}, Emily and {Menou}, Kristen},
        title = "{A General Circulation Model for Gaseous Exoplanets with Double-gray Radiative Transfer}",
      journal = {\apj},
         year = 2012,
        month = may,
       volume = {750},
       number = {2},
          eid = {96},
        pages = {96},
          doi = {10.1088/0004-637X/750/2/96},
archivePrefix = {arXiv},
       eprint = {1112.1658},
 primaryClass = {astro-ph.EP},
       adsurl = {https://ui.adsabs.harvard.edu/abs/2012ApJ...750...96R}
}

@Article{Mayne2014,
AUTHOR = {Mayne, N. J. and Baraffe, I. and Acreman, D. M. and Smith, C. and Wood, N. and Amundsen, D. S. and Thuburn, J. and Jackson, D. R.},
TITLE = {Using the UM dynamical cores to reproduce idealised 3-D flows},
JOURNAL = {Geoscientific Model Development},
VOLUME = {7},
YEAR = {2014},
NUMBER = {6},
PAGES = {3059--3087},
URL = {https://gmd.copernicus.org/articles/7/3059/2014/},
DOI = {10.5194/gmd-7-3059-2014}
}

@article{RauscherMenou2013,
	doi = {10.1088/0004-637x/764/1/103},
	year = 2013,
	month = {jan},
	publisher = {IOP Publishing},
	volume = {764},
	number = {1},
	pages = {103},
	author = {Emily Rauscher and Kristen Menou},
	title = {THREE-DIMENSIONAL ATMOSPHERIC CIRCULATION MODELS OF HD 189733b AND HD 209458b WITH CONSISTENT MAGNETIC DRAG AND OHMIC DISSIPATION},
	journal = {The Astrophysical Journal},

}

@ARTICLE{Christensen2009,
       author = {{Christensen}, Ulrich R. and {Holzwarth}, Volkmar and {Reiners}, Ansgar},
        title = "{Energy flux determines magnetic field strength of planets and stars}",
      journal = {\nat},
         year = 2009,
        month = jan,
       volume = {457},
       number = {7226},
        pages = {167-169},
          doi = {10.1038/nature07626},
       adsurl = {https://ui.adsabs.harvard.edu/abs/2009Natur.457..167C}
}

@ARTICLE{Birbky2018,
       author = {{Birkby}, J.~L.},
        title = "{Exoplanet Atmospheres at High Spectral Resolution}",
      journal = {arXiv e-prints},
         year = 2018,
        month = jun,
          eid = {arXiv:1806.04617},
        pages = {arXiv:1806.04617},
archivePrefix = {arXiv},
       eprint = {1806.04617},
 primaryClass = {astro-ph.EP},
       adsurl = {https://ui.adsabs.harvard.edu/abs/2018arXiv180604617B}
}

@ARTICLE{Snellen2010,
       author = {{Snellen}, Ignas A.~G. and {de Kok}, Remco J. and {de Mooij}, Ernst J.~W. and {Albrecht}, Simon},
        title = "{The orbital motion, absolute mass and high-altitude winds of exoplanet HD209458b}",
      journal = {\nat},
         year = 2010,
        month = jun,
       volume = {465},
       number = {7301},
        pages = {1049-1051},
          doi = {10.1038/nature09111},
archivePrefix = {arXiv},
       eprint = {1006.4364},
 primaryClass = {astro-ph.EP},
       adsurl = {https://ui.adsabs.harvard.edu/abs/2010Natur.465.1049S}
}

@ARTICLE{Brogi2019,
       author = {{Brogi}, Matteo and {Line}, Michael R.},
        title = "{Retrieving Temperatures and Abundances of Exoplanet Atmospheres with High-resolution Cross-correlation Spectroscopy}",
      journal = {\aj},
         year = 2019,
        month = mar,
       volume = {157},
       number = {3},
          eid = {114},
        pages = {114},
          doi = {10.3847/1538-3881/aaffd3},
archivePrefix = {arXiv},
       eprint = {1811.01681},
 primaryClass = {astro-ph.EP},
       adsurl = {https://ui.adsabs.harvard.edu/abs/2019AJ....157..114B}
}

@ARTICLE{Zhang2017,
       author = {{Zhang}, Jisheng and {Kempton}, Eliza M. -R. and {Rauscher}, Emily},
        title = "{Constraining Hot Jupiter Atmospheric Structure and Dynamics through Doppler-shifted Emission Spectra}",
      journal = {\apj},
         year = "2017",
        month = "Dec",
       volume = {851},
       number = {2},
          eid = {84},
        pages = {84},
          doi = {10.3847/1538-4357/aa9891},
archivePrefix = {arXiv},
       eprint = {1711.02684},
 primaryClass = {astro-ph.EP},
       adsurl = {https://ui.adsabs.harvard.edu/abs/2017ApJ...851...84Z}
}

@ARTICLE{Beltz2021,
       author = {{Beltz}, Hayley and {Rauscher}, Emily and {Brogi}, Matteo and {Kempton}, Eliza M. -R.},
        title = "{A Significant Increase in Detection of High-resolution Emission Spectra Using a Three-dimensional Atmospheric Model of a Hot Jupiter}",
      journal = {\aj},
         year = 2021,
        month = jan,
       volume = {161},
       number = {1},
          eid = {1},
        pages = {1},
          doi = {10.3847/1538-3881/abb67b},
archivePrefix = {arXiv},
       eprint = {2009.09030},
 primaryClass = {astro-ph.EP},
       adsurl = {https://ui.adsabs.harvard.edu/abs/2021AJ....161....1B}
}

@ARTICLE{Flowers2019,
       author = {{Flowers}, Erin and {Brogi}, Matteo and {Rauscher}, Emily and {Kempton}, Eliza M. -R. and {Chiavassa}, Andrea},
        title = "{The High-resolution Transmission Spectrum of HD 189733b Interpreted with Atmospheric Doppler Shifts from Three-dimensional General Circulation Models}",
      journal = {\aj},
         year = 2019,
        month = may,
       volume = {157},
       number = {5},
          eid = {209},
        pages = {209},
          doi = {10.3847/1538-3881/ab164c},
archivePrefix = {arXiv},
       eprint = {1810.06099},
 primaryClass = {astro-ph.EP},
       adsurl = {https://ui.adsabs.harvard.edu/abs/2019AJ....157..209F}
}

@ARTICLE{Lothringer2018,
       author = {{Lothringer}, Joshua D. and {Barman}, Travis and {Koskinen}, Tommi},
        title = "{Extremely Irradiated Hot Jupiters: Non-oxide Inversions, H$^{-}$ Opacity, and Thermal Dissociation of Molecules}",
      journal = {\apj},
         year = 2018,
        month = oct,
       volume = {866},
       number = {1},
          eid = {27},
        pages = {27},
          doi = {10.3847/1538-4357/aadd9e},
archivePrefix = {arXiv},
       eprint = {1805.00038},
 primaryClass = {astro-ph.EP},
       adsurl = {https://ui.adsabs.harvard.edu/abs/2018ApJ...866...27L}
}

@ARTICLE{Gibson2020,
       author = {{Gibson}, Neale P. and {Merritt}, Stephanie and {Nugroho}, Stevanus K. and {Cubillos}, Patricio E. and {de Mooij}, Ernst J.~W. and {Mikal-Evans}, Thomas and {Fossati}, Luca and {Lothringer}, Joshua and {Nikolov}, Nikolay and {Sing}, David K. and {Spake}, Jessica J. and {Watson}, Chris A. and {Wilson}, Jamie},
        title = "{Detection of Fe I in the atmosphere of the ultra-hot Jupiter WASP-121b, and a new likelihood-based approach for Doppler-resolved spectroscopy}",
      journal = {\mnras},
         year = 2020,
        month = apr,
       volume = {493},
       number = {2},
        pages = {2215-2228},
          doi = {10.1093/mnras/staa228},
archivePrefix = {arXiv},
       eprint = {2001.06430},
 primaryClass = {astro-ph.EP},
       adsurl = {https://ui.adsabs.harvard.edu/abs/2020MNRAS.493.2215G}
}

@article{Bell_2018,
	doi = {10.3847/2041-8213/aabcc8},
	url = {https://doi.org/10.3847%2F2041-8213%2Faabcc8},
	year = 2018,
	month = {apr},
	publisher = {American Astronomical Society},
	volume = {857},
	number = {2},
	pages = {L20},
	author = {Taylor J. Bell and Nicolas B. Cowan},
	title = {Increased Heat Transport in Ultra-hot Jupiter Atmospheres through H2 Dissociation and Recombination},
	journal = {The Astrophysical Journal},
}

@ARTICLE{parmentier2018,
       author = {{Parmentier}, Vivien and {Line}, Mike R. and {Bean}, Jacob L. and {Mansfield}, Megan and {Kreidberg}, Laura and {Lupu}, Roxana and {Visscher}, Channon and {D{\'e}sert}, Jean-Michel and {Fortney}, Jonathan J. and {Deleuil}, Magalie and {Arcangeli}, Jacob and {Showman}, Adam P. and {Marley}, Mark S.},
        title = "{From thermal dissociation to condensation in the atmospheres of ultra hot Jupiters: WASP-121b in context}",
      journal = {\aap},
         year = 2018,
        month = sep,
       volume = {617},
          eid = {A110},
        pages = {A110},
          doi = {10.1051/0004-6361/201833059},
archivePrefix = {arXiv},
       eprint = {1805.00096},
 primaryClass = {astro-ph.EP},
       adsurl = {https://ui.adsabs.harvard.edu/abs/2018A&A...617A.110P}
}

@article{simonnin2025time,
  title={Time-resolved absorption of six chemical species with MAROON-X points to a strong drag in the ultra-hot Jupiter TOI-1518 b},
  author={Simonnin, A and Parmentier, V and Wardenier, JP and Chauvin, G and Chiavassa, A and N’diaye, M and Tan, X and Heidari, N and Prinoth, B and Bean, J and others},
  journal={Astronomy \& Astrophysics},
  volume={698},
  pages={A314},
  year={2025},
  publisher={EDP Sciences}
}

@article{von2016inferring,
  title={Inferring asymmetric limb cloudiness on exoplanets from transit light curves},
  author={Von Paris, P and Gratier, P and Bord{\'e}, P and Leconte, J and Selsis, F},
  journal={Astronomy \& Astrophysics},
  volume={589},
  pages={A52},
  year={2016},
  publisher={EDP Sciences}
}

@article{chen2021evaluating,
  title={Evaluating large language models trained on code},
  author={Chen, Mark and Tworek, Jerry and Jun, Heewoo and Yuan, Qiming and Pinto, Henrique Ponde De Oliveira and Kaplan, Jared and Edwards, Harri and Burda, Yuri and Joseph, Nicholas and Brockman, Greg and others},
  journal={arXiv preprint arXiv:2107.03374},
  year={2021}
}

@article{dawson2016windspharm,
  title={Windspharm: A high-level library for global wind field computations using spherical harmonics},
  author={Dawson, Andrew},
  journal={Journal of Open Research Software},
  volume={4},
  number={1},
  year={2016}
}

@book{dutton2002ceaseless,
  title={The ceaseless wind: An introduction to the theory of atmospheric motion},
  author={Dutton, John A},
  year={2002},
  publisher={Courier Corporation}
}

@article{hammond2021rotational,
  title={The rotational and divergent components of atmospheric circulation on tidally locked planets},
  author={Hammond, Mark and Lewis, Neil T},
  journal={Proceedings of the National Academy of Sciences},
  volume={118},
  number={13},
  pages={e2022705118},
  year={2021},
  publisher={National Academy of Sciences}
}

@article{powell2019transit,
  title={Transit signatures of inhomogeneous clouds on hot Jupiters: insights from microphysical cloud modeling},
  author={Powell, Diana and Louden, Tom and Kreidberg, Laura and Zhang, Xi and Gao, Peter and Parmentier, Vivien},
  journal={The Astrophysical Journal},
  volume={887},
  number={2},
  pages={170},
  year={2019},
  publisher={The American Astronomical Society}
}

@article{kempton2017observational,
  title={An observational diagnostic for distinguishing between clouds and haze in hot exoplanet atmospheres},
  author={Kempton, Eliza M-R and Bean, Jacob L and Parmentier, Vivien},
  journal={The Astrophysical Journal Letters},
  volume={845},
  number={2},
  pages={L20},
  year={2017},
  publisher={The American Astronomical Society}
}

@ARTICLE{Yadav2017,
       author = {{Yadav}, Rakesh K. and {Thorngren}, Daniel P.},
        title = "{Estimating the Magnetic Field Strength in Hot Jupiters}",
      journal = {\apjl},
         year = 2017,
        month = nov,
       volume = {849},
       number = {1},
          eid = {L12},
        pages = {L12},
          doi = {10.3847/2041-8213/aa93fd},
archivePrefix = {arXiv},
       eprint = {1709.05676},
 primaryClass = {astro-ph.EP},
       adsurl = {https://ui.adsabs.harvard.edu/abs/2017ApJ...849L..12Y}
}

@article{Tan_2019,
	doi = {10.3847/1538-4357/ab4a76},
	url = {https://doi.org/10.3847/1538-4357/ab4a76},
	year = 2019,
	month = {nov},
	publisher = {American Astronomical Society},
	volume = {886},
	number = {1},
	pages = {26},
	author = {Xianyu Tan and Thaddeus D. Komacek},
	title = {The Atmospheric Circulation of Ultra-hot Jupiters},
	journal = {The Astrophysical Journal},

}

@ARTICLE{Christie2024damping,
       author = {{Christie}, D.~A. and {Mayne}, N.~J. and {Zamyatina}, M. and {Baskett}, H. and {Evans-Soma}, T.~M. and {Wood}, N. and {Kohary}, K.},
        title = "{Longitudinal filtering, sponge layers, and equatorial jet formation in a general circulation model of gaseous exoplanets}",
      journal = {\mnras},
         year = 2024,
        month = aug,
       volume = {532},
       number = {3},
        pages = {3001-3019},
          doi = {10.1093/mnras/stae1408},
archivePrefix = {arXiv},
       eprint = {2406.02231},
 primaryClass = {astro-ph.EP},
       adsurl = {https://ui.adsabs.harvard.edu/abs/2024MNRAS.532.3001C}
}

@ARTICLE{Deitrick2022THOR+HELIOS,
       author = {{Deitrick}, Russell and {Heng}, Kevin and {Schroffenegger}, Urs and {Kitzmann}, Daniel and {Grimm}, Simon L. and {Malik}, Matej and {Mendon{\c{c}}a}, Jo{\~a}o M. and {Morris}, Brett M.},
        title = "{The THOR + HELIOS general circulation model: multiwavelength radiative transfer with accurate scattering by clouds/hazes}",
      journal = {\mnras},
         year = 2022,
        month = may,
       volume = {512},
       number = {3},
        pages = {3759-3787},
          doi = {10.1093/mnras/stac680},
archivePrefix = {arXiv},
       eprint = {2203.02293},
 primaryClass = {astro-ph.EP},
       adsurl = {https://ui.adsabs.harvard.edu/abs/2022MNRAS.512.3759D}
}

@ARTICLE{TanShowman2020,
       author = {{Tan}, Xianyu and {Showman}, Adam P.},
        title = "{Atmospheric Circulation of Tidally Locked Gas Giants with Increasing Rotation and Implications for White Dwarf-Brown Dwarf Systems}",
      journal = {\apj},
         year = 2020,
        month = oct,
       volume = {902},
       number = {1},
          eid = {27},
        pages = {27},
          doi = {10.3847/1538-4357/abb3d4},
archivePrefix = {arXiv},
       eprint = {2001.06269},
 primaryClass = {astro-ph.EP},
       adsurl = {https://ui.adsabs.harvard.edu/abs/2020ApJ...902...27T}
}

@ARTICLE{Zhan2024bat,
       author = {{Zhan}, Ruizhi and {Koll}, Daniel D.~B. and {Ding}, Feng},
        title = "{Novel Atmospheric Dynamics Shape the Inner Edge of the Habitable Zone around White Dwarfs}",
      journal = {\apj},
         year = 2024,
        month = aug,
       volume = {971},
       number = {2},
          eid = {125},
        pages = {125},
          doi = {10.3847/1538-4357/ad54c1},
archivePrefix = {arXiv},
       eprint = {2406.03189},
 primaryClass = {astro-ph.EP},
       adsurl = {https://ui.adsabs.harvard.edu/abs/2024ApJ...971..125Z}
}

@ARTICLE{showman2013,
       author = {{Showman}, Adam P. and {Fortney}, Jonathan J. and {Lewis}, Nikole K. and {Shabram}, Megan},
        title = "{Doppler Signatures of the Atmospheric Circulation on Hot Jupiters}",
      journal = {\apj},
         year = 2013,
        month = jan,
       volume = {762},
       number = {1},
          eid = {24},
        pages = {24},
          doi = {10.1088/0004-637X/762/1/24},
archivePrefix = {arXiv},
       eprint = {1207.5639},
 primaryClass = {astro-ph.EP},
       adsurl = {https://ui.adsabs.harvard.edu/abs/2013ApJ...762...24S}
}

@ARTICLE{akin2025,
       author = {{Ak{\i}n}, C. and {Heng}, K. and {Mendon{\c{c}}a}, J.~M. and {Deitrick}, R. and {Gkouvelis}, L.},
        title = "{Global flow regimes of hot Jupiters}",
      journal = {\aap},
         year = 2025,
        month = jul,
       volume = {699},
          eid = {A74},
        pages = {A74},
          doi = {10.1051/0004-6361/202453597},
archivePrefix = {arXiv},
       eprint = {2505.12111},
 primaryClass = {astro-ph.EP},
       adsurl = {https://ui.adsabs.harvard.edu/abs/2025A&A...699A..74A}
}

@ARTICLE{Debras2020,
       author = {{Debras}, F. and {Mayne}, N. and {Baraffe}, I. and {Jaupart}, E. and {Mourier}, P. and {Laibe}, G. and {Goffrey}, T. and {Thuburn}, J.},
        title = "{Acceleration of superrotation in simulated hot Jupiter atmospheres}",
      journal = {\aap},
         year = 2020,
        month = jan,
       volume = {633},
          eid = {A2},
        pages = {A2},
          doi = {10.1051/0004-6361/201936110},
archivePrefix = {arXiv},
       eprint = {1911.03182},
 primaryClass = {astro-ph.EP},
       adsurl = {https://ui.adsabs.harvard.edu/abs/2020A&A...633A...2D}
}

@ARTICLE{PernaHeng2012,
       author = {{Perna}, Rosalba and {Heng}, Kevin and {Pont}, Fr{\'e}d{\'e}ric},
        title = "{The Effects of Irradiation on Hot Jovian Atmospheres: Heat Redistribution and Energy Dissipation}",
      journal = {\apj},
         year = 2012,
        month = may,
       volume = {751},
       number = {1},
          eid = {59},
        pages = {59},
          doi = {10.1088/0004-637X/751/1/59},
archivePrefix = {arXiv},
       eprint = {1201.5391},
 primaryClass = {astro-ph.EP},
       adsurl = {https://ui.adsabs.harvard.edu/abs/2012ApJ...751...59P}
}

@ARTICLE{Tsai2014,
       author = {{Tsai}, Shang-Min and {Dobbs-Dixon}, Ian and {Gu}, Pin-Gao},
        title = "{Three-dimensional Structures of Equatorial Waves and the Resulting Super-rotation in the Atmosphere of a Tidally Locked Hot Jupiter}",
      journal = {\apj},
         year = 2014,
        month = oct,
       volume = {793},
       number = {2},
          eid = {141},
        pages = {141},
          doi = {10.1088/0004-637X/793/2/141},
archivePrefix = {arXiv},
       eprint = {1405.0003},
 primaryClass = {astro-ph.EP},
       adsurl = {https://ui.adsabs.harvard.edu/abs/2014ApJ...793..141T}
}

@ARTICLE{Parmentier2015pf,
       author = {{Parmentier}, Vivien and {Guillot}, Tristan and {Fortney}, Jonathan J. and {Marley}, Mark S.},
        title = "{A non-grey analytical model for irradiated atmospheres. II. Analytical vs. numerical solutions}",
      journal = {\aap},
         year = 2015,
        month = feb,
       volume = {574},
          eid = {A35},
        pages = {A35},
          doi = {10.1051/0004-6361/201323127},
archivePrefix = {arXiv},
       eprint = {1311.6322},
 primaryClass = {astro-ph.EP},
       adsurl = {https://ui.adsabs.harvard.edu/abs/2015A&A...574A..35P}
}

@ARTICLE{Tan2024,
       author = {{Tan}, Xianyu and {Komacek}, Thaddeus D. and {Batalha}, Natasha E. and {Deming}, Drake and {Lupu}, Roxana and {Parmentier}, Vivien and {Pierrehumbert}, Raymond T.},
        title = "{Modelling the day-night temperature variations of ultra-hot Jupiters: confronting non-grey general circulation models and observations}",
      journal = {\mnras},
         year = 2024,
        month = feb,
       volume = {528},
       number = {1},
        pages = {1016-1036},
          doi = {10.1093/mnras/stae050},
archivePrefix = {arXiv},
       eprint = {2401.03859},
 primaryClass = {astro-ph.EP},
       adsurl = {https://ui.adsabs.harvard.edu/abs/2024MNRAS.528.1016T}
}

@ARTICLE{Hammond2022,
       author = {{Hammond}, Mark and {Abbot}, Dorian S.},
        title = "{Numerical dissipation strongly affects the equatorial jet speed in simulations of hot Jupiter atmospheres}",
      journal = {\mnras},
         year = 2022,
        month = apr,
       volume = {511},
       number = {2},
        pages = {2313-2325},
          doi = {10.1093/mnras/stac228},
       adsurl = {https://ui.adsabs.harvard.edu/abs/2022MNRAS.511.2313H}
}

@ARTICLE{Christie2025,
       author = {{Christie}, D.~A. and {Evans-Soma}, T.~M. and {Mayne}, N.~J. and {Kohary}, K.},
        title = "{Geometric considerations in hot Jupiter magnetic drag models}",
      journal = {\mnras},
         year = 2025,
        month = aug,
       volume = {541},
       number = {3},
        pages = {2773-2789},
          doi = {10.1093/mnras/staf1146},
archivePrefix = {arXiv},
       eprint = {2507.08511},
 primaryClass = {astro-ph.EP},
       adsurl = {https://ui.adsabs.harvard.edu/abs/2025MNRAS.541.2773C}
}

@ARTICLE{Blocker2026,
       author = {{Bl{\"o}cker}, Aljona and {Carone}, Ludmila and {Helling}, Christiane},
        title = "{Inhomogeneous magnetic coupling in exoplanets: the stop \& go of WASP-18 b's atmospheric flows}",
      journal = {arXiv e-prints},
         year = 2026,
        month = feb,
          eid = {arXiv:2602.18101},
        pages = {arXiv:2602.18101},
          doi = {10.48550/arXiv.2602.18101},
archivePrefix = {arXiv},
       eprint = {2602.18101},
 primaryClass = {astro-ph.EP},
       adsurl = {https://ui.adsabs.harvard.edu/abs/2026arXiv260218101B}
}

@ARTICLE{Rothgrid2024,
       author = {{Roth}, Alexander and {Parmentier}, Vivien and {Hammond}, Mark},
        title = "{Hot Jupiter diversity and the onset of TiO/VO revealed by a large grid of non-grey global circulation models}",
      journal = {\mnras},
         year = 2024,
        month = jun,
       volume = {531},
       number = {1},
        pages = {1056-1083},
          doi = {10.1093/mnras/stae984},
archivePrefix = {arXiv},
       eprint = {2404.09626},
 primaryClass = {astro-ph.EP},
       adsurl = {https://ui.adsabs.harvard.edu/abs/2024MNRAS.531.1056R}
}

@ARTICLE{Pelletier2023,
       author = {{Pelletier}, Stefan and {Benneke}, Bj{\"o}rn and {Ali-Dib}, Mohamad and {Prinoth}, Bibiana and {Kasper}, David and {Seifahrt}, Andreas and {Bean}, Jacob L. and {Debras}, Florian and {Klein}, Baptiste and {Bazinet}, Luc and {Hoeijmakers}, H. Jens and {Kesseli}, Aurora Y. and {Lim}, Olivia and {Carmona}, Andres and {Pino}, Lorenzo and {Casasayas-Barris}, N{\'u}ria and {Hood}, Thea and {St{\"u}rmer}, Julian},
        title = "{Vanadium oxide and a sharp onset of cold-trapping on a giant exoplanet}",
      journal = {\nat},
         year = 2023,
        month = jul,
       volume = {619},
       number = {7970},
        pages = {491-494},
          doi = {10.1038/s41586-023-06134-0},
archivePrefix = {arXiv},
       eprint = {2306.08739},
 primaryClass = {astro-ph.EP},
       adsurl = {https://ui.adsabs.harvard.edu/abs/2023Natur.619..491P}
}

@ARTICLE{Simonnin2025,
       author = {{Simonnin}, A. and {Parmentier}, V. and {Wardenier}, J.~P. and {Chauvin}, G. and {Chiavassa}, A. and {N'Diaye}, M. and {Tan}, X. and {Heidari}, N. and {Prinoth}, B. and {Bean}, J. and {H{\'e}brard}, G. and {Line}, M. and {Kitzmann}, D. and {Kasper}, D. and {Pelletier}, S. and {Seidel}, J.~V. and {Seifhart}, A. and {Benneke}, B. and {Bonfils}, X. and {Brogi}, M. and {D{\'e}sert}, J.-M. and {Gandhi}, S. and {Hammond}, M. and {Lee}, E.~K.~H. and {Moutou}, C. and {Palma-Bifani}, P. and {Pino}, L. and {Rauscher}, E. and {Weiner Mansfield}, M. and {Serrano Bell}, J. and {Smith}, P.},
        title = "{Time-resolved absorption of six chemical species with MAROON-X points to a strong drag in the ultra-hot Jupiter TOI-1518 b}",
      journal = {\aap},
         year = 2025,
        month = jun,
       volume = {698},
          eid = {A314},
        pages = {A314},
          doi = {10.1051/0004-6361/202453241},
archivePrefix = {arXiv},
       eprint = {2412.01472},
 primaryClass = {astro-ph.EP},
       adsurl = {https://ui.adsabs.harvard.edu/abs/2025A&A...698A.314S}
}

@ARTICLE{Komacek2025hysteresis,
       author = {{Komacek}, Thaddeus D.},
        title = "{Limited Hysteresis in the Atmospheric Dynamics of Hot Jupiters}",
      journal = {\apj},
         year = 2025,
        month = apr,
       volume = {983},
       number = {1},
          eid = {7},
        pages = {7},
          doi = {10.3847/1538-4357/adbae9},
archivePrefix = {arXiv},
       eprint = {2502.19394},
 primaryClass = {astro-ph.EP},
       adsurl = {https://ui.adsabs.harvard.edu/abs/2025ApJ...983....7K}
}

@ARTICLE{Perna2010magdrag,
       author = {{Perna}, Rosalba and {Menou}, Kristen and {Rauscher}, Emily},
        title = "{{\GG{20020530}}Magnetic Drag on Hot Jupiter Atmospheric Winds}",
      journal = {\apj},
         year = 2010,
        month = aug,
       volume = {719},
       number = {2},
        pages = {1421-1426},
          doi = {10.1088/0004-637X/719/2/1421},
archivePrefix = {arXiv},
       eprint = {1003.3838},
 primaryClass = {astro-ph.EP},
       adsurl = {https://ui.adsabs.harvard.edu/abs/2010ApJ...719.1421P}
}

@ARTICLE{Ehrenreich2020,
       author = {{Ehrenreich}, David and {Lovis}, Christophe and {Allart}, Romain and {Zapatero Osorio}, Mar{\'\i}a Rosa and {Pepe}, Francesco and {Cristiani}, Stefano and {Rebolo}, Rafael and {Santos}, Nuno C. and {Borsa}, Francesco and {Demangeon}, Olivier and {Dumusque}, Xavier and {Gonz{\'a}lez Hern{\'a}ndez}, Jonay I. and {Casasayas-Barris}, N{\'u}ria and {S{\'e}gransan}, Damien and {Sousa}, S{\'e}rgio and {Abreu}, Manuel and {Adibekyan}, Vardan and {Affolter}, Michael and {Allende Prieto}, Carlos and {Alibert}, Yann and {Aliverti}, Matteo and {Alves}, David and {Amate}, Manuel and {Avila}, Gerardo and {Baldini}, Veronica and {Bandy}, Timothy and {Benz}, Willy and {Bianco}, Andrea and {Bolmont}, {\'E}meline and {Bouchy}, Fran{\c{c}}ois and {Bourrier}, Vincent and {Broeg}, Christopher and {Cabral}, Alexandre and {Calderone}, Giorgio and {Pall{\'e}}, Enric and {Cegla}, H.~M. and {Cirami}, Roberto and {Coelho}, Jo{\~a}o M.~P. and {Conconi}, Paolo and {Coretti}, Igor and {Cumani}, Claudio and {Cupani}, Guido and {Dekker}, Hans and {Delabre}, Bernard and {Deiries}, Sebastian and {D'Odorico}, Valentina and {Di Marcantonio}, Paolo and {Figueira}, Pedro and {Fragoso}, Ana and {Genolet}, Ludovic and {Genoni}, Matteo and {G{\'e}nova Santos}, Ricardo and {Hara}, Nathan and {Hughes}, Ian and {Iwert}, Olaf and {Kerber}, Florian and {Knudstrup}, Jens and {Landoni}, Marco and {Lavie}, Baptiste and {Lizon}, Jean-Louis and {Lendl}, Monika and {Lo Curto}, Gaspare and {Maire}, Charles and {Manescau}, Antonio and {Martins}, C.~J.~A.~P. and {M{\'e}gevand}, Denis and {Mehner}, Andrea and {Micela}, Giusi and {Modigliani}, Andrea and {Molaro}, Paolo and {Monteiro}, Manuel and {Monteiro}, Mario and {Moschetti}, Manuele and {M{\"u}ller}, Eric and {Nunes}, Nelson and {Oggioni}, Luca and {Oliveira}, Ant{\'o}nio and {Pariani}, Giorgio and {Pasquini}, Luca and {Poretti}, Ennio and {Rasilla}, Jos{\'e} Luis and {Redaelli}, Edoardo and {Riva}, Marco and {Santana Tschudi}, Samuel and {Santin}, Paolo and {Santos}, Pedro and {Segovia Milla}, Alex and {Seidel}, Julia V. and {Sosnowska}, Danuta and {Sozzetti}, Alessandro and {Span{\`o}}, Paolo and {Su{\'a}rez Mascare{\~n}o}, Alejandro and {Tabernero}, Hugo and {Tenegi}, Fabio and {Udry}, St{\'e}phane and {Zanutta}, Alessio and {Zerbi}, Filippo},
        title = "{Nightside condensation of iron in an ultrahot giant exoplanet}",
      journal = {\nat},
         year = 2020,
        month = apr,
       volume = {580},
       number = {7805},
        pages = {597-601},
          doi = {10.1038/s41586-020-2107-1},
archivePrefix = {arXiv},
       eprint = {2003.05528},
 primaryClass = {astro-ph.EP},
       adsurl = {https://ui.adsabs.harvard.edu/abs/2020Natur.580..597E}
}

@INPROCEEDINGS{Maroonx2018,
       author = {{Seifahrt}, Andreas and {St{\"u}rmer}, Julian and {Bean}, Jacob L. and {Schwab}, Christian},
        title = "{MAROON-X: a radial velocity spectrograph for the Gemini Observatory}",
    booktitle = {Ground-based and Airborne Instrumentation for Astronomy VII},
         year = 2018,
       editor = {{Evans}, Christopher J. and {Simard}, Luc and {Takami}, Hideki},
       series = {Society of Photo-Optical Instrumentation Engineers (SPIE) Conference Series},
       volume = {10702},
        month = jul,
          eid = {107026D},
        pages = {107026D},
          doi = {10.1117/12.2312936},
archivePrefix = {arXiv},
       eprint = {1805.09276},
 primaryClass = {astro-ph.IM},
       adsurl = {https://ui.adsabs.harvard.edu/abs/2018SPIE10702E..6DS}
}

@ARTICLE{NIRPS2025,
       author = {{Bouchy}, Fran{\c{c}}ois and {Doyon}, Ren{\'e} and {Pepe}, Francesco and {Melo}, Claudio and {Artigau}, {\'E}tienne and {Malo}, Lison and {Wildi}, Fran{\c{c}}ois and {Baron}, Fr{\'e}d{\'e}rique and {Delfosse}, Xavier and {De Medeiros}, Jose Renan and {Rebolo}, Rafael and {Santos}, Nuno C. and {Wade}, Gregg and {Allart}, Romain and {Al Moulla}, Khaled and {Blind}, Nicolas and {Cadieux}, Charles and {Canto Martins}, Bruno L. and {Cook}, Neil J. and {Dumusque}, Xavier and {Frensch}, Yolanda and {Genest}, Fr{\'e}d{\'e}ric and {Gonz{\'a}lez Hern{\'a}ndez}, Jonay I. and {Grieves}, Nolan and {Lo Curto}, Gaspare and {Lovis}, Christophe and {Mignon}, Lucile and {Nielsen}, Louise D. and {Poulin-Girard}, Anne-Sophie and {Rasilla}, Jos{\'e} Luis and {Reshetov}, Vladimir and {Sosnowska}, Danuta and {Sordet}, Michael and {Saint-Antoine}, Jonathan and {Su{\'a}rez Mascare{\~n}o}, Alejandro and {Thibault}, Simon and {Vall{\'e}e}, Philippe and {Vandal}, Thomas and {Abreu}, Manuel and {Aguiar}, Jos{\'e} L.~A. and {Allain}, Guillaume and {Arial}, Tomy and {Auger}, Hugues and {Barros}, Susana C.~C. and {Bazinet}, Luc and {Benneke}, Bj{\"o}rn and {Bonfils}, Xavier and {Boucher}, Anne and {Bourrier}, Vincent and {Bovay}, S{\'e}bastien and {Broeg}, Christopher and {Brousseau}, Denis and {Bruniquel}, Vincent and {Bryan}, Marta and {Cabral}, Alexandre and {Carmona}, Andres and {Carteret}, Yann and {Challita}, Zalpha and {Chazelas}, Bruno and {Cloutier}, Ryan and {Coelho}, Jo{\~a}o and {Cointepas}, Marion and {Conod}, Uriel and {Cowan}, Nicolas B. and {Cristo}, Eduardo and {Gomes da Silva}, Jo{\~a}o and {Dauplaise}, Laurie and {Darveau-Bernier}, Antoine and {de Lima Gomes}, Roseane and {de Freitas}, Daniel Brito and {Delgado-Mena}, Elisa and {Delisle}, Jean-Baptiste and {Ehrenreich}, David and {Faria}, Jo{\~a}o and {Figueira}, Pedro and {Fontinele}, Dasaev O. and {Forveille}, Thierry and {Gagn{\'e}}, Jonathan and {Genolet}, Ludovic and {T{\'e}mich}, F{\'e}lix Gracia and {Hernandez}, Olivier and {Hobson}, Melissa J. and {Hoeijmakers}, Jens and {Hubin}, Norbert and {Jahandar}, Farbod and {Jayawardhana}, Ray and {K{\"a}ufl}, Hans-Ulrich and {Kerley}, Dan and {Kolb}, Johann and {Krishnamurthy}, Vigneshwaran and {Lafreni{\`e}re}, David and {Lamontagne}, Pierrot and {Larue}, Pierre and {Leath}, Henry and {L'Heureux}, Alexandrine and {de Castro Le{\~a}o}, Izan and {Lim}, Olivia and {Martins}, Allan M. and {Matthews}, Jaymie and {Mayer}, Jean-S{\'e}bastien and {Messias}, Yuri S. and {Metchev}, Stan and {Moranta}, Leslie and {Mordasini}, Christoph and {Mounzer}, Dany and {Nari}, Nicola and {Osborn}, Ares and {Ouellet}, Mathieu and {Otegi}, Jon and {Parc}, L{\'e}na and {Pasquini}, Luca and {Passegger}, Vera M. and {Pelletier}, Stefan and {Peroux}, C{\'e}line and {Piaulet-Ghorayeb}, Caroline and {Plotnykov}, Mykhaylo and {Pompei}, Emanuela and {Rowe}, Jason and {Sarajlic}, Mirsad and {Segovia}, Alex and {Seidel}, Julia and {S{\'e}gransan}, Damien and {Schnell}, Robin and {Costa Silva}, Ana Rita and {Srivastava}, Avidaan and {Stefanov}, Atanas K. and {Teixeira}, M{\'a}rcio A. and {Udry}, St{\'e}phane and {Valencia}, Diana and {Vaulato}, Valentina and {Wardenier}, Joost P. and {Wehbe}, Bachar and {Weisserman}, Drew and {Wevers}, Ivan and {Yariv}, Vincent and {Zins}, G{\'e}rard},
        title = "{NIRPS joining HARPS at ESO 3.6 m: On-sky performance and science objectives}",
      journal = {\aap},
         year = 2025,
        month = aug,
       volume = {700},
          eid = {A10},
        pages = {A10},
          doi = {10.1051/0004-6361/202453341},
archivePrefix = {arXiv},
       eprint = {2507.21767},
 primaryClass = {astro-ph.IM},
       adsurl = {https://ui.adsabs.harvard.edu/abs/2025A&A...700A..10B}
}

@ARTICLE{RauscherKempton2014,
       author = {{Rauscher}, Emily and {Kempton}, Eliza M.~R.},
        title = "{The Atmospheric Circulation and Observable Properties of Non-synchronously Rotating Hot Jupiters}",
      journal = {\apj},
         year = 2014,
        month = jul,
       volume = {790},
       number = {1},
          eid = {79},
        pages = {79},
          doi = {10.1088/0004-637X/790/1/79},
archivePrefix = {arXiv},
       eprint = {1402.4833},
 primaryClass = {astro-ph.EP},
       adsurl = {https://ui.adsabs.harvard.edu/abs/2014ApJ...790...79R}
}

@ARTICLE{ZhangPulsar,
       author = {{Zhang}, Michael and {Beleznay}, Maya and {Brandt}, Timothy D. and {Romani}, Roger W. and {Gao}, Peter and {Beltz}, Hayley and {Bailes}, Matthew and {Nixon}, Matthew C. and {Bean}, Jacob L. and {Komacek}, Thaddeus D. and {Coy}, Brandon P. and {Fu}, Guangwei and {Luque}, Rafael and {Reardon}, Daniel J. and {Carli}, Emma and {Shannon}, Ryan M. and {Fortney}, Jonathan J. and {Piette}, Anjali A.~A. and {Miller}, M. Coleman and {Desert}, Jean-Michel},
        title = "{A Carbon-rich Atmosphere on a Windy Pulsar Planet}",
      journal = {\apjl},
         year = 2025,
        month = dec,
       volume = {995},
       number = {2},
          eid = {L64},
        pages = {L64},
          doi = {10.3847/2041-8213/ae157c},
archivePrefix = {arXiv},
       eprint = {2509.04558},
 primaryClass = {astro-ph.EP},
       adsurl = {https://ui.adsabs.harvard.edu/abs/2025ApJ...995L..64Z}
}

@ARTICLE{RauscherMenou20212,
       author = {{Rauscher}, Emily and {Menou}, Kristen},
        title = "{A General Circulation Model for Gaseous Exoplanets with Double-gray Radiative Transfer}",
      journal = {\apj},
         year = 2012,
        month = may,
       volume = {750},
       number = {2},
          eid = {96},
        pages = {96},
          doi = {10.1088/0004-637X/750/2/96},
archivePrefix = {arXiv},
       eprint = {1112.1658},
 primaryClass = {astro-ph.EP},
       adsurl = {https://ui.adsabs.harvard.edu/abs/2012ApJ...750...96R}
}

@ARTICLE{Pepsi2025,
       author = {{Strassmeier}, K.~G. and {Ilyin}, I. and {J{\"a}rvinen}, A. and {Weber}, M. and {Woche}, M. and {Barnes}, S.~I. and {Bauer}, S.-M. and {Beckert}, E. and {Bittner}, W. and {Bredthauer}, R. and {Carroll}, T.~A. and {Denker}, C. and {Dionies}, F. and {DiVarano}, I. and {D{\"o}scher}, D. and {Fechner}, T. and {Feuerstein}, D. and {Granzer}, T. and {Hahn}, T. and {Harnisch}, G. and {Hofmann}, A. and {Lesser}, M. and {Paschke}, J. and {Pankratow}, S. and {Plank}, V. and {Pl{\"u}schke}, D. and {Popow}, E. and {Sablowski}, D.},
        title = "{PEPSI: The high-resolution {\'e}chelle spectrograph and polarimeter for the Large Binocular Telescope}",
      journal = {Astronomische Nachrichten},
         year = 2015,
        month = may,
       volume = {336},
       number = {4},
        pages = {324},
          doi = {10.1002/asna.201512172},
archivePrefix = {arXiv},
       eprint = {1505.06492},
 primaryClass = {astro-ph.IM},
       adsurl = {https://ui.adsabs.harvard.edu/abs/2015AN....336..324S}
}

@ARTICLE{Komacek2020,
       author = {{Komacek}, Thaddeus D. and {Showman}, Adam P.},
        title = "{Temporal Variability in Hot Jupiter Atmospheres}",
      journal = {\apj},
         year = 2020,
        month = jan,
       volume = {888},
       number = {1},
          eid = {2},
        pages = {2},
          doi = {10.3847/1538-4357/ab5b0b},
archivePrefix = {arXiv},
       eprint = {1910.09523},
 primaryClass = {astro-ph.EP},
       adsurl = {https://ui.adsabs.harvard.edu/abs/2020ApJ...888....2K}
}

@ARTICLE{Zhan2024batrotator,
       author = {{Zhan}, Ruizhi and {Koll}, Daniel D.~B. and {Ding}, Feng},
        title = "{Novel Atmospheric Dynamics Shape the Inner Edge of the Habitable Zone around White Dwarfs}",
      journal = {\apj},
         year = 2024,
        month = aug,
       volume = {971},
       number = {2},
          eid = {125},
        pages = {125},
          doi = {10.3847/1538-4357/ad54c1},
archivePrefix = {arXiv},
       eprint = {2406.03189},
 primaryClass = {astro-ph.EP},
       adsurl = {https://ui.adsabs.harvard.edu/abs/2024ApJ...971..125Z}
}

@ARTICLE{Komacek2022,
       author = {{Komacek}, Thaddeus D. and {Tan}, Xianyu and {Gao}, Peter and {Lee}, Elspeth K.~H.},
        title = "{Patchy Nightside Clouds on Ultra-hot Jupiters: General Circulation Model Simulations with Radiatively Active Cloud Tracers}",
      journal = {\apj},
         year = 2022,
        month = jul,
       volume = {934},
       number = {1},
          eid = {79},
        pages = {79},
          doi = {10.3847/1538-4357/ac7723},
archivePrefix = {arXiv},
       eprint = {2205.07834},
 primaryClass = {astro-ph.EP},
       adsurl = {https://ui.adsabs.harvard.edu/abs/2022ApJ...934...79K}
}

@ARTICLE{Wardenier2021,
       author = {{Wardenier}, Joost P. and {Parmentier}, Vivien and {Lee}, Elspeth K.~H. and {Line}, Michael R. and {Gharib-Nezhad}, Ehsan},
        title = "{Decomposing the iron cross-correlation signal of the ultra-hot Jupiter WASP-76b in transmission using 3D Monte Carlo radiative transfer}",
      journal = {\mnras},
         year = 2021,
        month = sep,
       volume = {506},
       number = {1},
        pages = {1258-1283},
          doi = {10.1093/mnras/stab1797},
archivePrefix = {arXiv},
       eprint = {2105.11034},
 primaryClass = {astro-ph.EP},
       adsurl = {https://ui.adsabs.harvard.edu/abs/2021MNRAS.506.1258W}
}

@ARTICLE{Wardenier2022,
       author = {{Wardenier}, Joost P. and {Parmentier}, Vivien and {Lee}, Elspeth K.~H.},
        title = "{All along the line of sight: a closer look at opening angles and absorption regions in the atmospheres of transiting exoplanets}",
      journal = {\mnras},
         year = 2022,
        month = feb,
       volume = {510},
       number = {1},
        pages = {620-629},
          doi = {10.1093/mnras/stab3432},
archivePrefix = {arXiv},
       eprint = {2111.11830},
 primaryClass = {astro-ph.EP},
       adsurl = {https://ui.adsabs.harvard.edu/abs/2022MNRAS.510..620W}
}

@ARTICLE{Seidel2025,
       author = {{Seidel}, Julia V. and {Prinoth}, Bibiana and {Pino}, Lorenzo and {dos Santos}, Leonardo A. and {Chakraborty}, Hritam and {Parmentier}, Vivien and {Sedaghati}, Elyar and {Wardenier}, Joost P. and {Farret Jentink}, Casper and {Rosa Zapatero Osorio}, Maria and {Allart}, Romain and {Ehrenreich}, David and {Lendl}, Monika and {Roccetti}, Giulia and {Damasceno}, Yuri and {Bourrier}, Vincent and {Lillo-Box}, Jorge and {Hoeijmakers}, H. Jens and {Pall{\'e}}, Enric and {Santos}, Nuno and {Su{\'a}rez Mascare{\~n}o}, Alejandro and {Sousa}, Sergio G. and {Tabernero}, Hugo M. and {Pepe}, Francesco A.},
        title = "{Vertical structure of an exoplanet's atmospheric jet stream}",
      journal = {arXiv e-prints},
         year = 2025,
        month = feb,
          eid = {arXiv:2502.12261},
        pages = {arXiv:2502.12261},
          doi = {10.48550/arXiv.2502.12261},
archivePrefix = {arXiv},
       eprint = {2502.12261},
 primaryClass = {astro-ph.EP},
       adsurl = {https://ui.adsabs.harvard.edu/abs/2025arXiv250212261S}
}

@ARTICLE{Espressopepe2014,
       author = {{Pepe}, F. and {Molaro}, P. and {Cristiani}, S. and {Rebolo}, R. and {Santos}, N.~C. and {Dekker}, H. and {M{\'e}gevand}, D. and {Zerbi}, F.~M. and {Cabral}, A. and {Di Marcantonio}, P. and {Abreu}, M. and {Affolter}, M. and {Aliverti}, M. and {Allende Prieto}, C. and {Amate}, M. and {Avila}, G. and {Baldini}, V. and {Bristow}, P. and {Broeg}, C. and {Cirami}, R. and {Coelho}, J. and {Conconi}, P. and {Coretti}, I. and {Cupani}, G. and {D'Odorico}, V. and {De Caprio}, V. and {Delabre}, B. and {Dorn}, R. and {Figueira}, P. and {Fragoso}, A. and {Galeotta}, S. and {Genolet}, L. and {Gomes}, R. and {Gonz{\'a}lez Hern{\'a}ndez}, J.~I. and {Hughes}, I. and {Iwert}, O. and {Kerber}, F. and {Landoni}, M. and {Lizon}, J. -L. and {Lovis}, C. and {Maire}, C. and {Mannetta}, M. and {Martins}, C. and {Monteiro}, M. and {Oliveira}, A. and {Poretti}, E. and {Rasilla}, J.~L. and {Riva}, M. and {Santana Tschudi}, S. and {Santos}, P. and {Sosnowska}, D. and {Sousa}, S. and {Span{\'o}}, P. and {Tenegi}, F. and {Toso}, G. and {Vanzella}, E. and {Viel}, M. and {Zapatero Osorio}, M.~R.},
        title = "{ESPRESSO: The next European exoplanet hunter}",
      journal = {Astronomische Nachrichten},
         year = 2014,
        month = jan,
       volume = {335},
       number = {1},
        pages = {8},
          doi = {12.1002/asna.201312004},
       adsurl = {https://ui.adsabs.harvard.edu/abs/2014AN....335....8P}
}

@ARTICLE{Snellen2025,
       author = {{Snellen}, Ignas A.~G.},
        title = "{Exoplanet Atmospheres at High Spectral Resolution}",
      journal = {\araa},
         year = 2025,
        month = aug,
       volume = {63},
       number = {1},
        pages = {83-125},
          doi = {10.1146/annurev-astro-052622-031342},
archivePrefix = {arXiv},
       eprint = {2505.08926},
 primaryClass = {astro-ph.EP},
       adsurl = {https://ui.adsabs.harvard.edu/abs/2025ARA&A..63...83S}
}

@ARTICLE{Wardenier2024,
       author = {{Wardenier}, Joost P. and {Parmentier}, Vivien and {Line}, Michael R. and {Weiner Mansfield}, Megan and {Tan}, Xianyu and {Tsai}, Shang-Min and {Bean}, Jacob L. and {Birkby}, Jayne L. and {Brogi}, Matteo and {D{\'e}sert}, Jean-Michel and {Gandhi}, Siddharth and {Lee}, Elspeth K.~H. and {Levens}, Colette I. and {Pino}, Lorenzo and {Smith}, Peter C.~B.},
        title = "{Phase-resolving the Absorption Signatures of Water and Carbon Monoxide in the Atmosphere of the Ultra-hot Jupiter WASP-121b with GEMINI-S/IGRINS}",
      journal = {\pasp},
         year = 2024,
        month = aug,
       volume = {136},
       number = {8},
          eid = {084403},
        pages = {084403},
          doi = {10.1088/1538-3873/ad5c9f},
archivePrefix = {arXiv},
       eprint = {2406.09641},
 primaryClass = {astro-ph.EP},
       adsurl = {https://ui.adsabs.harvard.edu/abs/2024PASP..136h4403W}
}

@ARTICLE{Prinoth2023,
       author = {{Prinoth}, B. and {Hoeijmakers}, H.~J. and {Pelletier}, S. and {Kitzmann}, D. and {Morris}, B.~M. and {Seifahrt}, A. and {Kasper}, D. and {Korhonen}, H.~H. and {Burheim}, M. and {Bean}, J.~L. and {Benneke}, B. and {Borsato}, N.~W. and {Brady}, M. and {Grimm}, S.~L. and {Luque}, R. and {St{\"u}rmer}, J. and {Thorsbro}, B.},
        title = "{Time-resolved transmission spectroscopy of the ultra-hot Jupiter WASP-189 b}",
      journal = {\aap},
         year = 2023,
        month = oct,
       volume = {678},
          eid = {A182},
        pages = {A182},
          doi = {10.1051/0004-6361/202347262},
archivePrefix = {arXiv},
       eprint = {2308.04523},
 primaryClass = {astro-ph.EP},
       adsurl = {https://ui.adsabs.harvard.edu/abs/2023A&A...678A.182P}
}

@ARTICLE{KesseliBeltz2024,
       author = {{Kesseli}, Aurora Y. and {Beltz}, Hayley and {Rauscher}, Emily and {Snellen}, I.~A.~G.},
        title = "{Up, Up, and Away: Winds and Dynamical Structure as a Function of Altitude in the Ultrahot Jupiter WASP-76b}",
      journal = {\apj},
         year = 2024,
        month = nov,
       volume = {975},
       number = {1},
          eid = {9},
        pages = {9},
          doi = {10.3847/1538-4357/ad772f},
archivePrefix = {arXiv},
       eprint = {2409.03124},
 primaryClass = {astro-ph.EP},
       adsurl = {https://ui.adsabs.harvard.edu/abs/2024ApJ...975....9K}
}

@ARTICLE{2022Savel,
       author = {{Savel}, Arjun B. and {Kempton}, Eliza M. -R. and {Malik}, Matej and {Komacek}, Thaddeus D. and {Bean}, Jacob L. and {May}, Erin M. and {Stevenson}, Kevin B. and {Mansfield}, Megan and {Rauscher}, Emily},
        title = "{No Umbrella Needed: Confronting the Hypothesis of Iron Rain on WASP-76b with Post-processed General Circulation Models}",
      journal = {\apj},
         year = 2022,
        month = feb,
       volume = {926},
       number = {1},
          eid = {85},
        pages = {85},
          doi = {10.3847/1538-4357/ac423f},
archivePrefix = {arXiv},
       eprint = {2109.00163},
 primaryClass = {astro-ph.EP},
       adsurl = {https://ui.adsabs.harvard.edu/abs/2022ApJ...926...85S}
}

@ARTICLE{Pino2022,
       author = {{Pino}, L. and {Brogi}, M. and {D{\'e}sert}, J.~M. and {Nascimbeni}, V. and {Bonomo}, A.~S. and {Rauscher}, E. and {Basilicata}, M. and {Biazzo}, K. and {Bignamini}, A. and {Borsa}, F. and {Claudi}, R. and {Covino}, E. and {Di Mauro}, M.~P. and {Guilluy}, G. and {Maggio}, A. and {Malavolta}, L. and {Micela}, G. and {Molinari}, E. and {Molinaro}, M. and {Montalto}, M. and {Nardiello}, D. and {Pedani}, M. and {Piotto}, G. and {Poretti}, E. and {Rainer}, M. and {Scandariato}, G. and {Sicilia}, D. and {Sozzetti}, A.},
        title = "{The GAPS Programme at TNG. XLI. The climate of KELT-9b revealed with a new approach to high-spectral-resolution phase curves}",
      journal = {\aap},
         year = 2022,
        month = dec,
       volume = {668},
          eid = {A176},
        pages = {A176},
          doi = {10.1051/0004-6361/202244593}
}

@ARTICLE{Borsa2021,
       author = {{Borsa}, F. and {Allart}, R. and {Casasayas-Barris}, N. and {Tabernero}, H. and {Zapatero Osorio}, M.~R. and {Cristiani}, S. and {Pepe}, F. and {Rebolo}, R. and {Santos}, N.~C. and {Adibekyan}, V. and {Bourrier}, V. and {Demangeon}, O.~D.~S. and {Ehrenreich}, D. and {Pall{\'e}}, E. and {Sousa}, S. and {Lillo-Box}, J. and {Lovis}, C. and {Micela}, G. and {Oshagh}, M. and {Poretti}, E. and {Sozzetti}, A. and {Allende Prieto}, C. and {Alibert}, Y. and {Amate}, M. and {Benz}, W. and {Bouchy}, F. and {Cabral}, A. and {Dekker}, H. and {D'Odorico}, V. and {Di Marcantonio}, P. and {Figueira}, P. and {Genova Santos}, R. and {Gonz{\'a}lez Hern{\'a}ndez}, J.~I. and {Lo Curto}, G. and {Manescau}, A. and {Martins}, C.~J.~A.~P. and {M{\'e}gevand}, D. and {Mehner}, A. and {Molaro}, P. and {Nunes}, N.~J. and {Riva}, M. and {Su{\'a}rez Mascare{\~n}o}, A. and {Udry}, S. and {Zerbi}, F.},
        title = "{Atmospheric Rossiter-McLaughlin effect and transmission spectroscopy of WASP-121b with ESPRESSO}",
      journal = {\aap},
         year = 2021,
        month = jan,
       volume = {645},
          eid = {A24},
        pages = {A24},
          doi = {10.1051/0004-6361/202039344},
archivePrefix = {arXiv},
       eprint = {2011.01245},
 primaryClass = {astro-ph.EP},
       adsurl = {https://ui.adsabs.harvard.edu/abs/2021A&A...645A..24B}
}

@article{Roman_2021,
	doi = {10.3847/1538-4357/abd549},
	url = {https://doi.org/10.3847/1538-4357/abd549},
	year = 2021,
	month = {feb},
	publisher = {American Astronomical Society},
	volume = {908},
	number = {1},
	pages = {101},
	author = {Michael T. Roman and Eliza M.-R. Kempton and Emily Rauscher and Caleb K. Harada and Jacob L. Bean and Kevin B. Stevenson},
	title = {Clouds in Three-dimensional Models of Hot Jupiters over a Wide Range of Temperatures. I. Thermal Structures and Broadband Phase-curve Predictions},
	journal = {The Astrophysical Journal},
}

@ARTICLE{Rogers2017,
       author = {{Rogers}, T.~M.},
        title = "{Constraints on the magnetic field strength of HAT-P-7 b and other hot giant exoplanets}",
      journal = {Nature Astronomy},
         year = 2017,
        month = jun,
       volume = {1},
          eid = {0131},
        pages = {0131},
          doi = {10.1038/s41550-017-0131},
archivePrefix = {arXiv},
       eprint = {1704.06271},
 primaryClass = {astro-ph.EP},
       adsurl = {https://ui.adsabs.harvard.edu/abs/2017NatAs...1E.131R}
}

@article{Rogers_2014b,
	doi = {10.1088/0004-637x/794/2/132},
	url = {https://doi.org/10.1088/0004-637x/794/2/132},
	year = 2014,
	month = {oct},
	publisher = {American Astronomical Society},
	volume = {794},
	number = {2},
	pages = {132},
	author = {T. M. Rogers and T. D. Komacek},
	title = {{MAGNETIC} {EFFECTS} {IN} {HOT} {JUPITER} {ATMOSPHERES}},
	journal = {The Astrophysical Journal},
}

@INPROCEEDINGS{CRIRES+2014,
       author = {{Follert}, R. and {Dorn}, R.~J. and {Oliva}, E. and {Lizon}, J.~L. and {Hatzes}, A. and {Piskunov}, N. and {Reiners}, A. and {Seemann}, U. and {Stempels}, E. and {Heiter}, U. and {Marquart}, T. and {Lockhart}, M. and {Anglada-Escude}, G. and {L{\"o}winger}, T. and {Baade}, D. and {Grunhut}, J. and {Bristow}, P. and {Klein}, B. and {Jung}, Y. and {Ives}, D.~J. and {Kerber}, F. and {Pozna}, E. and {Paufique}, J. and {Kaeufl}, H.~U. and {Origlia}, L. and {Valenti}, E. and {Gojak}, D. and {Hilker}, M. and {Pasquini}, L. and {Smette}, A. and {Smoker}, J.},
        title = "{CRIRES+: a cross-dispersed high-resolution infrared spectrograph for the ESO VLT}",
    booktitle = {Ground-based and Airborne Instrumentation for Astronomy V},
         year = 2014,
       editor = {{Ramsay}, Suzanne K. and {McLean}, Ian S. and {Takami}, Hideki},
       series = {Society of Photo-Optical Instrumentation Engineers (SPIE) Conference Series},
       volume = {9147},
        month = jul,
          eid = {914719},
        pages = {914719},
          doi = {10.1117/12.2054197},
       adsurl = {https://ui.adsabs.harvard.edu/abs/2014SPIE.9147E..19F}
}

@ARTICLE{PerezBecker2013,
       author = {{Perez-Becker}, Daniel and {Showman}, Adam P.},
        title = "{Atmospheric Heat Redistribution on Hot Jupiters}",
      journal = {\apj},
         year = 2013,
        month = oct,
       volume = {776},
       number = {2},
          eid = {134},
        pages = {134},
          doi = {10.1088/0004-637X/776/2/134},
archivePrefix = {arXiv},
       eprint = {1306.4673},
 primaryClass = {astro-ph.EP},
       adsurl = {https://ui.adsabs.harvard.edu/abs/2013ApJ...776..134P}
}

@ARTICLE{Guillot2010,
       author = {{Guillot}, T.},
        title = "{On the radiative equilibrium of irradiated planetary atmospheres}",
      journal = {\aap},
         year = 2010,
        month = sep,
       volume = {520},
          eid = {A27},
        pages = {A27},
          doi = {10.1051/0004-6361/200913396},
archivePrefix = {arXiv},
       eprint = {1006.4702},
 primaryClass = {astro-ph.EP},
       adsurl = {https://ui.adsabs.harvard.edu/abs/2010A&A...520A..27G}
}

@ARTICLE{Showman2002,
       author = {{Showman}, A.~P. and {Guillot}, T.},
        title = "{Atmospheric circulation and tides of ``51 Pegasus b-like'' planets}",
      journal = {\aap},
         year = 2002,
        month = apr,
       volume = {385},
        pages = {166-180},
          doi = {10.1051/0004-6361:20020101},
archivePrefix = {arXiv},
       eprint = {astro-ph/0202236},
 primaryClass = {astro-ph},
       adsurl = {https://ui.adsabs.harvard.edu/abs/2002A&A...385..166S}
}

@ARTICLE{2019Arcangeli,
       author = {{Arcangeli}, Jacob and {D{\'e}sert}, Jean-Michel and {Parmentier}, Vivien and {Stevenson}, Kevin B. and {Bean}, Jacob L. and {Line}, Michael R. and {Kreidberg}, Laura and {Fortney}, Jonathan J. and {Showman}, Adam P.},
        title = "{Climate of an ultra hot Jupiter. Spectroscopic phase curve of WASP-18b with HST/WFC3}",
      journal = {\aap},
         year = 2019,
        month = may,
       volume = {625},
          eid = {A136},
        pages = {A136},
          doi = {10.1051/0004-6361/201834891},
archivePrefix = {arXiv},
       eprint = {1904.02069},
 primaryClass = {astro-ph.EP},
       adsurl = {https://ui.adsabs.harvard.edu/abs/2019A&A...625A.136A}
}

@ARTICLE{2011ShomanPolvani,
       author = {{Showman}, Adam P. and {Polvani}, Lorenzo M.},
        title = "{Equatorial Superrotation on Tidally Locked Exoplanets}",
      journal = {\apj},
         year = 2011,
        month = sep,
       volume = {738},
       number = {1},
          eid = {71},
        pages = {71},
          doi = {10.1088/0004-637X/738/1/71},
archivePrefix = {arXiv},
       eprint = {1103.3101},
 primaryClass = {astro-ph.EP},
       adsurl = {https://ui.adsabs.harvard.edu/abs/2011ApJ...738...71S}
}

@INPROCEEDINGS{IGRINS2014,
       author = {{Park}, Chan and {Jaffe}, Daniel T. and {Yuk}, In-Soo and {Chun}, Moo-Young and {Pak}, Soojong and {Kim}, Kang-Min and {Pavel}, Michael and {Lee}, Hanshin and {Oh}, Heeyoung and {Jeong}, Ueejeong and {Sim}, Chae Kyung and {Lee}, Hye-In and {Nguyen Le}, Huynh Anh and {Strubhar}, Joseph and {Gully-Santiago}, Michael and {Oh}, Jae Sok and {Cha}, Sang-Mok and {Moon}, Bongkon and {Park}, Kwijong and {Brooks}, Cynthia and {Ko}, Kyeongyeon and {Han}, Jeong-Yeol and {Nah}, Jakyoung and {Hill}, Peter C. and {Lee}, Sungho and {Barnes}, Stuart and {Yu}, Young Sam and {Kaplan}, Kyle and {Mace}, Gregory and {Kim}, Hwihyun and {Lee}, Jae-Joon and {Hwang}, Narae and {Park}, Byeong-Gon},
        title = "{Design and early performance of IGRINS (Immersion Grating Infrared Spectrometer)}",
    booktitle = {Ground-based and Airborne Instrumentation for Astronomy V},
         year = 2014,
       editor = {{Ramsay}, Suzanne K. and {McLean}, Ian S. and {Takami}, Hideki},
       series = {Society of Photo-Optical Instrumentation Engineers (SPIE) Conference Series},
       volume = {9147},
        month = jul,
          eid = {91471D},
        pages = {91471D},
          doi = {10.1117/12.2056431},
       adsurl = {https://ui.adsabs.harvard.edu/abs/2014SPIE.9147E..1DP}
}

@ARTICLE{2018Drummond,
       author = {{Drummond}, B. and {Mayne}, N.~J. and {Manners}, J. and {Carter}, A.~L. and {Boutle}, I.~A. and {Baraffe}, I. and {H{\'e}brard}, {\'E}. and {Tremblin}, P. and {Sing}, D.~K. and {Amundsen}, D.~S. and {Acreman}, D.},
        title = "{Observable Signatures of Wind-driven Chemistry with a Fully Consistent Three-dimensional Radiative Hydrodynamics Model of HD 209458b}",
      journal = {\apjl},
         year = 2018,
        month = mar,
       volume = {855},
       number = {2},
          eid = {L31},
        pages = {L31},
          doi = {10.3847/2041-8213/aab209},
archivePrefix = {arXiv},
       eprint = {1802.09222},
 primaryClass = {astro-ph.EP},
       adsurl = {https://ui.adsabs.harvard.edu/abs/2018ApJ...855L..31D}
}

@ARTICLE{2011ShowmanPolvani,
       author = {{Showman}, Adam P. and {Polvani}, Lorenzo M.},
        title = "{Equatorial Superrotation on Tidally Locked Exoplanets}",
      journal = {\apj},
         year = 2011,
        month = sep,
       volume = {738},
       number = {1},
          eid = {71},
        pages = {71},
          doi = {10.1088/0004-637X/738/1/71},
archivePrefix = {arXiv},
       eprint = {1103.3101},
 primaryClass = {astro-ph.EP},
       adsurl = {https://ui.adsabs.harvard.edu/abs/2011ApJ...738...71S}
}

@ARTICLE{Malsky2024PF,
       author = {{Malsky}, Isaac and {Rauscher}, Emily and {Roman}, Michael T. and {Lee}, Elspeth K.~H. and {Beltz}, Hayley and {Savel}, Arjun and {Kempton}, Eliza M. -R. and {Cinque}, L.},
        title = "{A Direct Comparison between the Use of Double Gray and Multiwavelength Radiative Transfer in a General Circulation Model with and without Radiatively Active Clouds}",
      journal = {\apj},
         year = 2024,
        month = jan,
       volume = {961},
       number = {1},
          eid = {66},
        pages = {66},
          doi = {10.3847/1538-4357/ad0b70},
archivePrefix = {arXiv},
       eprint = {2311.01506},
 primaryClass = {astro-ph.EP},
       adsurl = {https://ui.adsabs.harvard.edu/abs/2024ApJ...961...66M}
}

@article{kurucz1995kurucz,
  title={Kurucz CD-Rom No. 23},
  author={Kurucz, RL},
  journal={Atomic Line List},
  year={1995},
  publisher={Harvard-Smithsonian Center for Astrophysics}
}
\bibliographystyle{aasjournal}



\end{document}